\documentclass[twocolumn,tighten]{aastex631}

\newcommand{\nustar}{\textit{NuSTAR}}

\usepackage{relsize}
\usepackage{url}
\accepted{May 15, 2024}
\published{July 29, 2024}
\submitjournal{ApJ}
\reportnum{2024 ApJ 970, 167}
\graphicspath{{./}{figures/}}

\begin{document}

\title{A \nustar\ Census of the X-ray Binary Population of the M31 Disk}

\author[0009-0004-3597-6975]{Hannah Moon}
\affiliation{Department of Physics \& Astronomy, The University of Utah, 115 South 1400 East, Salt Lake City, UT 84112, USA}

\author[0000-0001-9110-2245]{Daniel R. Wik}
\affiliation{Department of Physics \& Astronomy, The University of Utah, 115 South 1400 East, Salt Lake City, UT 84112, USA}

\author[0000-0001-7539-1593]{V. Antoniou}
\affiliation{Department of Physics \& Astronomy, Texas Tech University, Box 41051, Lubbock, TX 79409-1051, USA
}
\affiliation{Center for Astrophysics $\vert$ Harvard \& Smithsonian, 60 Garden Street, Cambridge MA 02138, USA}

\author[0000-0002-3719-940X]{M. Eracleous}
\affiliation{Department of Astronomy and Astrophysics and Institute for Gravitation and the Cosmos, The Pennsylvania State University, 525 Davey Lab, University Park, PA 16802}

\author[0000-0001-8667-2681]{Ann E. Hornschemeier}
\affiliation{NASA Goddard Space Flight Center, Code 662, Greenbelt, MD 20771, USA}
\affiliation{The Johns Hopkins University, Homewood Campus, Baltimore, MD 21218, USA}

\author[0000-0003-3252-352X]{Margaret Lazzarini}
\affiliation{Department of Physics \& Astronomy, California State University Los Angeles, 5151 State University Drive, Los Angeles, CA 90032, USA}

\author[0000-0003-2192-3296]{Bret D. Lehmer}
\affiliation{ Department of Physics, University of Arkansas, 226 Physics Building, 825 West Dickson Street, Fayetteville, AR 72701, USA}

\author[0000-0001-7855-8336]{Neven Vulic}
\affiliation{Eureka Scientific, Inc., 2452 Delmer St., Suite 100, Oakland, CA 94602-3017, USA}
\affiliation{NASA Goddard Space Flight Center, Code 662, Greenbelt, MD 20771, USA}
\affiliation{Department of Astronomy, University of Maryland, College Park, MD 20742-2421, USA}
\affiliation{Center for Research and Exploration in Space Science and Technology, NASA/GSFC, Greenbelt, MD 20771, USA}

\author[0000-0002-7502-0597]{Benjamin F. Williams}
\affiliation{Department of Astronomy, Box 351580, University of Washington, Seattle, WA 98195, USA}

\author[0000-0003-0976-4755]{T. J. Maccarone}
\affiliation{Department of Physics \& Astronomy, Texas Tech University, Box 41051, Lubbock, TX 79409-1051, USA
}

\author[0000-0002-4656-6881]{K. Pottschmidt}
\affiliation{Astroparticle Physics Laboratory, NASA Goddard Space Flight Center, Greenbelt, MD 20771, USA}
\affiliation{CRESST, Center for Space Sciences and Technology, UMBC, Baltimore, MD 210250, USA}

\author[0000-0001-5655-1440]{Andrew Ptak}
\affiliation{NASA Goddard Space Flight Center, Code 662, Greenbelt, MD 20771, USA}
\affiliation{The Johns Hopkins University, Homewood Campus, Baltimore, MD 21218, USA}

\author[0000-0001-6366-3459]{Mihoko Yukita}
\affiliation{NASA Goddard Space Flight Center, Code 662, Greenbelt, MD 20771, USA}
\affiliation{The Johns Hopkins University, Homewood Campus, Baltimore, MD 21218, USA}

\author[0000-0001-8952-676X]{Andreas Zezas}
\affiliation{Physics Department, University of Crete, Heraklion, Greece}
\affiliation{Center for Astrophysics $\vert$ Harvard \& Smithsonian, 60 Garden Street, Cambridge MA 02138, USA}
\affiliation{ Institute of Astrophysics, Foundation for Research and Technology-Hellas, N. Plastira 100, Vassilika Vouton, 71110 Heraklion, Greece}

\begin{abstract}

Using hard (E$>$10~keV) X-ray observations with \textit{NuSTAR}, we are able to differentiate between the accretion states, and thus compact object types, of neutron stars and black holes in X-ray binaries (XRBs) in M31, our nearest Milky Way-type neighbor. 
Using ten moderate-depth (20-50 ks) observations of the disk of M31 covering a total of $\sim$0.45 deg$^2$, we detect 20 sources at 2$\sigma$ in the 4--25~keV band pass, 14 of which we consider to be XRB candidates. This complements an existing, deeper (100-400 ks) survey covering $\sim$0.2 deg$^2$ of the bulge and the northeastern disk. 
We make tentative classifications of 9 of these sources with the use of diagnostic color-intensity and color-color diagrams, which separate sources into various neutron star and black hole regimes, identifying 3 black holes and 6 neutron stars. In addition, we create X-ray luminosity functions for both the full (4--25~keV) and hard (12--25~keV) band, as well as sub-populations of the full band based on compact object type and association with globular clusters. 
Our best fit globular cluster XLF is shallower than the field XLF, and preliminary BH and NS XLFs suggest a difference in shape based on compact object type.  
We find that the cumulative disk XLFs in the full and hard band are best fit by power laws with indices of $1.32$ and $1.28$ respectively.  
This is consistent with models of the Milky Way XLF from \cite{Grimm02}, \cite{Voss10}, and \cite{Doroshenko14}.

\end{abstract}


\section{Introduction} 
\label{sec:intro}
The X-ray binary population of a galaxy can provide insights into many aspects of the history, makeup, and structure of that galaxy. 
As our nearest Milky Way-type neighbor, the Andromeda galaxy, or Messier 31 (M31), is an excellent laboratory for learning about our own Galaxy. 

The X-ray Luminosity Function (XLF) of a galaxy can be linked to its star formation history and current star formation rate \citep{Kilgard02,Grimm03,Prestwich09}, as well as to stellar age and mass \citep{Gilfanov04, Lehmer14}. 
In addition, studies of accreting black hole and neutron star populations can provide insights into the history of star formation and evolution in a galaxy through binary population synthesis modeling.

Thanks to its high-energy ($E>10$~keV) sensitivity, \nustar\ is uniquely well-suited for characterizing the hard band XLF and constraining the nature of the compact objects in X-ray binary (XRB) systems.
XRB population demographics in a galaxy are incredibly useful for constraining key parameters in binary evolution models, which are necessary to predict the production of gravitational wave (GW) sources \citep{Giacobbo18}.
The success of these models relies on accurate initial parameters that describe isolated binary systems like those found in $L^*$ galaxies \citep{Schechter76} like the Milky Way and M31.
Increasing the number of fully classified XRBs in a variety of environments will allow for a better determination of  these initial parameters.  

Past work on XRB population demographics with \nustar\ has focused on nearby (d$<$30 Mpc) bright star-forming galaxies such as NGC 253, M81, M82, M33, and M83 \citep{Wik14a, Lehmer15, Yukita16, Vulic18, Yang22}. 
In these cases, the proximity of the galaxies allowed for resolution of individual sources, which in turn allowed for characterization of the compact objects in XRBs to determine BH and NS populations.
In \cite{Lazzarini18}, \nustar\ observations of the inner disk region of M31 were paired with \textit{Chandra} data and \textit{Hubble Space Telescope} observations from the PHAT survey to fully characterize the compact object and stellar companion in 15 candidate HMXBs.
In the case of M33, this analysis was paired with PHATTER observations to fully characterize 7 HMXBs in the same manner \citep{Lazzarini23}.

Other than the Milky Way, M31 is the only  $L^*$-type galaxy 
where this type of work is feasible with \nustar . 
While we can detect lower luminosity sources in the Milky Way, uncertainties in the distances to sources and large and varying absorption columns in the plane of the Galaxy complicate completeness estimates of population studies.
M31 is located at an ideal distance where the distance uncertainty is significantly reduced, but we can still resolve individual sources with \nustar.
This has already proven useful for compact object characterization in the central region of M31.
In addition, we can use results from studies of M31 to inform and compare with results from similar studies in the Milky Way.

\begin{figure*}
\centering  
\includegraphics[scale=0.33]{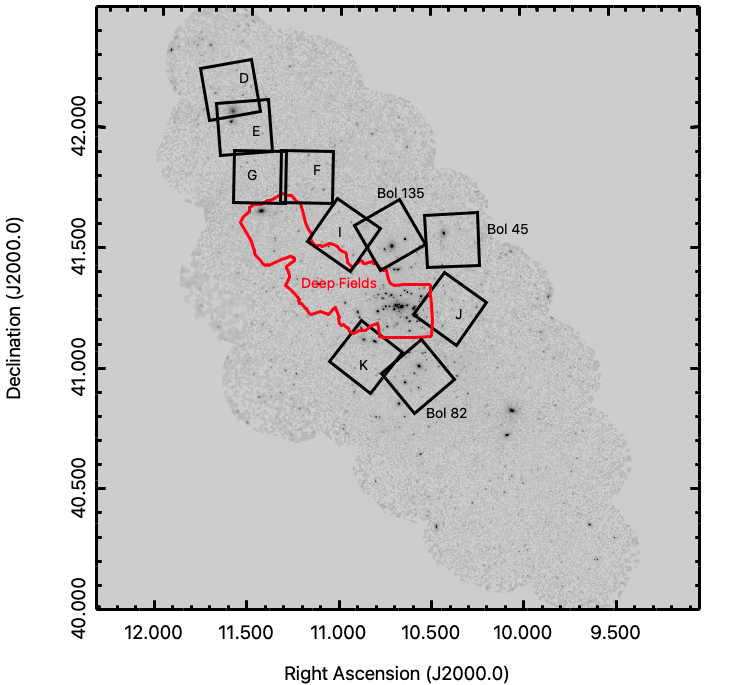}
\includegraphics[scale=0.32]{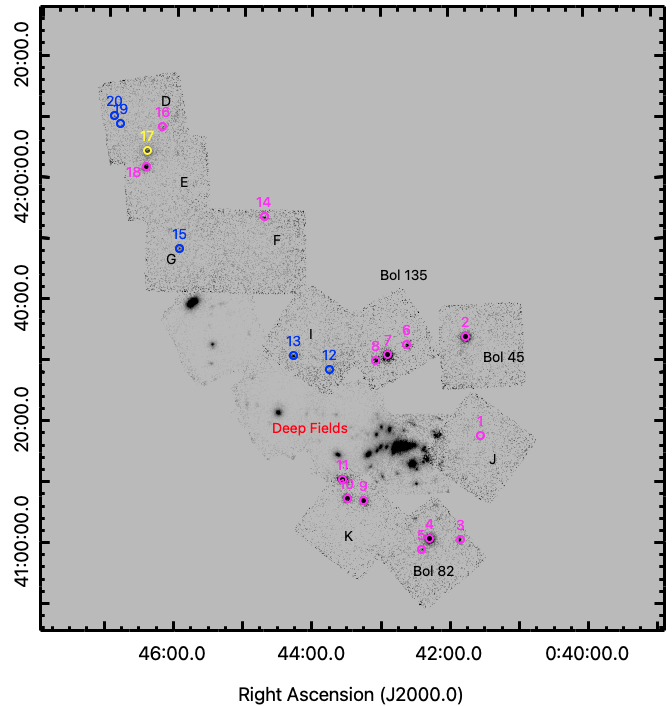}
\caption{Existing \nustar\ coverage of M31. Left: Smoothed, square root scaled 2--7.2~keV 
{\it XMM-Newton} mosaic of the full M31 disk, overlaid with a footprint of all existing \nustar\ fields. Fields analyzed in this paper are outlined in black; prior deep ($\sim$400~ks) field observations are in red. Right: Background-subtracted and exposure corrected 4--25~keV mosaic of the current coverage of M31 with \nustar. The \nustar\ mosaic is asinh-scaled and smoothed. 
The numbered sources and their locations are shown, with XRB candidates in magenta, background galaxies in blue, and the galaxy cluster in yellow.
RA and Dec are shown in degrees. \label{fig:foot}}
\end{figure*}

Here, we analyze data from ten fields observed by \textit{NuSTAR} with $\sim$40~ks of cleaned exposure time per field. These fields, shown in Figure \ref{fig:foot} overlaid on an {\it XMM-Newton} mosaic of the galaxy\footnote{\url{https://heasarc.gsfc.nasa.gov/docs/xmm/gallery/esas-gallery/xmm_gal_science_m31.html}}, cover large portions of the disk of M31. We construct color-color and color-intensity diagrams for all of the sources in our sample and use these as diagnostics to broadly classify compact objects in the XRBs. 
In addition, we create full and hard band X-ray luminosity functions for our fields, combined with deep fields analyzed in Moon et al., in prep, which we refer to as the Deep Paper going forward.

 \begin{deluxetable*}{ccccc}
\tablecaption{Observations \label{tab:obs}}
\tablewidth{0pt}
\tablehead{
\colhead{UT Start Date}& \colhead{Field Name} &
\colhead{Observation ID} & \colhead{Original Exposure (ks)} &
\colhead{Cleaned\tablenotemark{a} Exposure (A,B) (ks)}
}

\startdata
2017 Nov 07 & Field D & 50312001002 & 47.8 & 41.1, 39.4 \\
2017 Nov 11 & Field E & 50312002002 & 51.8 & 44.7, 42.9 \\
2017 Nov 13 & Field F & 50312003002 & 49.0 & 41.3, 42.4\\
2017 Nov 14 & Field G & 50312004002 & 48.8 & 45.0, 45.5\\
2018 Dec 19 & Field I & 50460002002 & 48.4 & 40.0, 36.8\\
2018 Dec 20 & Field J & 50460003002 & 48.5 & 37.7, 38.8\\
2018 Dec 24 & Field K & 50460004002 & 47.7 & 44.3, 41.9 \\
2017 Aug 04 & BOL 135 & 30365002002 & 21.7 & 18.2, 17.9 \\
2018 Jan 12  & BOL 82 & 30365001002 & 42.4 & 39.4, 37.7 \\
2019 Nov 11 & BOL 45 & 30560002002 & 42.9 & 38.3, 38.9 \\
\enddata
\tablenotetext{a}{Cleaned exposure time refers to the total exposure time after eliminating periods of high background as described in Section \ref{sec:reduction}.}
 \end{deluxetable*}

\section{Data and initial reduction} \label{sec:reduction}

We make use of ten archival \textit{NuSTAR} observations of M31, representing 10 distinct fields spanning much of the northern disk. Each observation represents $\sim$40ks of exposure time (Table~\ref{tab:obs}).
 
We choose to filter the light curves of these datasets by hand to maximize the total good time intervals (GTIs) and remove bad background periods. Excluding obvious sources, we extract light curves from each observation in 100 second time bins. 
A first light curve is made with data in the range $50~{\rm keV}<E<160$~keV to identify high background periods caused by high energy particles and resulting radiation.
These periods are identified by eye based on their deviation from the average distribution of rates.
Typically, there is a roughly day-long oscillation that varies in intensity as well as an orbital variation on a $\sim$95~min timescale.
Otherwise, the light curve is quite stable, making any large anomalies higher than the baseline and easy to identify. 
These high background periods are typically caused by passage through the South Atlantic Anomaly (SAA), solar flaring, and other periods of higher background.
We repeat this process in the 1.6--20~keV band, which allows us to remove periods of high background due to solar flaring.
This procedure typically removes a few ks of high background from each observation.

For background subtraction, we follow the procedure outlined in \cite{Wik14a} and the associated code, {\tt nuskybgd}, which uses an empirical model to simultaneously fit the background spectra extracted from source-free regions in the field of view. 
The background models account for four possible sources of background contribution: internal background, aperture stray light from the cosmic X-ray background (CXB), scattered and reflected stray light from the Sun, and focused emission from the CXB.
The internal background comes from the radiation environment of \nustar's orbit, as well as various activation and fluorescence lines.
The aperture stray light models the fraction of unfocused stray light that gets past the aperture stops and is proportional to the solid angle visible through the aperture stops.
Scattered and reflected stray light models the background contributions from solar stray light that is reflected from various parts of the telescope (i.e., potentially the back of the aperture stops, although the origin is not yet understood) and into the detector.
Finally, the focused cosmic background models the background contribution of the CXB within the field of view.

We create four box regions, about 6$^\prime$ on a side, to be coincident with the four detectors in each focal plane, from which to extract background spectra.
These spectra are extracted, and corresponding models are constructed with \texttt{nuskybgd} and fit to the spectra to find the normalization of each background component. We use the best-fit background parameters to produce background and background-subtracted images and spectra for all of the energy bands used in this work: 4--6 keV, 6--12 keV, 12--25 keV, and 4--25 keV. A smoothed, background-subtracted mosaic of the 10 fields in this work and the deeper observations from the Deep Paper are shown in Figure \ref{fig:foot}. 
 
\section{Catalog Creation}
\label{sec:catalog}

In order to create a catalog of possible sources, we visually identify candidate sources in the combined, smoothed, and background-subtracted images in the 4--25~keV band. Square regions are created around these candidate sources as the extraction areas of energy band images for point-spread function (PSF) fitting. In instances where multiple sources may have overlapping PSFs, they are contained within the same square or rectangular region. Model PSFs are created for each observation and source, centered on the source coordinates identified by eye, plus any additional shifts necessary to ensure that the models and data are accurately registered to each other.
Detailed examples of this procedure, where multiple sources are in close angular proximity to each other, were first developed and described in \cite{Wik14b} and can also be seen in the Deep Paper.
In this work, source overlap is much less severe, but we use the same method, which is outlined below.

In the fitting procedure, each point source and background model for each observation are forward-modeled and fit to the image raw event data, with only their normalizations as a free parameter.
This fitting is performed by minimizing the C-statistic \citep{cash}, and the uncertainties are estimated by adjusting the parameter values and refitting until the change in the C-statistic corresponds to the desired probability; since the C-statistic is scaled to follow the $\chi^{2}$ distribution, we use these $\Delta C$ changes to define probability uncertainty ranges (e.g., $\Delta C = 1.0$ providing 68\% or 1$\sigma$ uncertainty ranges).

For each dataset, a raw image is extracted from the event file, filtered by the rectangular region and the energy band under consideration.
A model background image is similarly made with {\tt nuskybgd} from the parameters of the fit to the background, also done with {\tt nuskybgd}.
Unvignetted exposure maps are also created for the region, which account for lost sensitivity due to chip gaps and dead pixels.
These maps are multiplied by the source models, allowing the response of the detectors to be reflected in fits to the data.
For each point source, a composite PSF model is created given the energy band (taking into account changes in the PSF shape with energy), weighted assuming power law ($\Gamma=2$) emission inside the band; the PSF model only very weakly depends on this choice, as the PSF shape varies only slightly with energy, especially for narrow bandpasses.

To fit the data image, the background model image---scaled by a free parameter---is added to the PSF model(s) of the sources in the region, each of which is also scaled by free parameters normalized to return the total count rate of the source in that band.
The data and summed model images are both binned so that there is at least 1 count per bin, to allow the C-statistic to be used to find the best fit model parameters.
To find the best fit, the normalization parameters on the models are varied until C-stat is minimized.
The uncertainties are then derived by varying each parameter until $\Delta C$ changes by the desired amount to achieve the confidence interval for one interesting parameter (as described above).
In the case when multiple energy bands are fit simultaneously (see next section), the procedure is the same except that model parameters are related across bands and include hardness ratios, from which count rates in individual bands can be derived.

Candidate sources are added or removed until all candidate sources have count rates with $>1\sigma$ significance in a given band. We complete this procedure in the 4--6~keV, 6--12~keV, 12--25~keV, and 4--25~keV bands, hereby referred to as soft (S), medium (M), hard (H), and full (F) bands, respectively. We then eliminate any sources from the resulting list that are not detected with $>2\sigma$ significance in the 4--25~keV band.
The $>2\sigma$ detection threshold, necessary to extract useful information from the data in narrower bandpasses, also makes automated/unbiased source detection unnecessary, as by-eye image inspection easily picks out point-source-like features at lower significance than this.
We remove any sources classified as AGN by either the \cite{Stiele11}, \cite{Barnard14}, or \cite{Williams18} catalogs, removing five known AGN.
Sources 17a and 17b are the same source in two observations, and known to be a galaxy cluster \citep{Kotov06}.
After removing these known AGN, we identify 14 candidate XRB sources across all fields. 

\subsection{Creation of Hardness Diagnostics}

With the goal of identifying the compact object and accretion state of each XRB source, we perform the fitting procedure described above to the three narrow energy bands simultaneously in order to constrain the hardness ratios. 
Though we do not attempt to classify the known AGN, we do calculate and include their hardness ratios.
For 19 out of our 20 sources, we were able to do this fitting following the same procedure that we used to fit each of the narrow bands individually as described earlier in Section \ref{sec:catalog}. 
The simultaneous fitting uses two hardness ratios and the full band (4--25~keV) count rate as free parameters, computing the individual count rates in each of our bands from those, along with the $1\sigma$ errors.
Our two hardness ratios are defined as
\begin{equation}
    h_{1} = \frac{(M-S)}{(M+S)} \text{  and  } h_{2}=\frac{(H-M)}{(H+M)}
\end{equation}
and the full band rate is $F = S+M+H$. 
The narrow band rates can then be calculated using:

\begin{eqnarray}
    S =& \frac{C\cdot F}{1+C+D}\, ,&\\
    M =& \frac{F}{1+C+D}\, ,&~{\rm and}\\
    H =& \frac{D\cdot F}{1+C+D}\, ,&
\end{eqnarray}

where

\begin{equation}
    C=\frac{1-h_{1}}{1+h_{1}}\text{ and }D=\frac{1+h_{2}}{1-h_{2}}\, .
\end{equation}

By using the hardness ratios directly in simultaneous fits to the three energy band images, uncertainty ranges can be directly computed for those quantities using all the information available at once.

For the remaining source that we were unable to fit due to the low number of counts (Source~1 in Table~\ref{tab:rates}), we perform this process by hand. 
We collected the total counts in each narrow band using a $30^{\prime\prime}$ radius region centered on the source. 
We do this separately for each sub band image created after our data calibration without background subtraction, keeping the A and B detectors separate. 
We then use the same regions to extract the number of background counts for each band from each background image produced from the background modeling, and combine the background and background-subtracted count rates from each detector. 

The hardness ratios are defined as above, and the one sigma errors on hardness ratios based on low energy band $i$ and high energy band $i+1$ are calculated following propagation of errors, which results in the expression
\begin{equation}
    \sigma_{h} = 2\frac{\sqrt{(r_{i}\sigma_{i+1})^{2} + (r_{i+1}\sigma_{i})^{2}}}{(r_{i}+r_{i+1})^2}\, ,
\end{equation}
 where $r_{i}$ and $\sigma_{i}$ are the count rates and corresponding errors in energy band $i$.

\subsection{Flux and Luminosity Conversions}

Converting count rates and hardness ratios to incident fluxes and intrinsic luminosities requires a spectral model.
For simplicity, we assume that our spectra can be reasonably well fit by a single power law at hard ($E>6$~keV) energies.
We create model power law spectra with different power law indices $\Gamma$, incorporating \nustar\ response files and the {\tt XSpec} \texttt{fakeit} procedure.
We use these model spectra to obtain count rates, from which we calculate the hardness ratio $h_2$ as a function of input $\Gamma$, and fit an analytic function that relates our hardness ratio $h_{2}$ to the power law index using {\tt scipy}'s \texttt{curve\_fit} function. 

\begin{deluxetable}{c|c|c|c}[ht]
 \label{tab:params}
\tabletypesize{\scriptsize}
\tablecaption{Equation \ref{eq:flux} Parameters}
\tablewidth{0pt}
\tablehead{\colhead{Equation} & \colhead{$X$} &\colhead{$Y$} &\colhead{$Z$}}
\startdata
Flux$_{4-25~keV}$ & $5.4091$ & $5.5349\times10^{-1}$ & $2.1685$ \\
Flux$_{12-25~keV}$ & $12.547$ & $6.6413\times10^{-2}$ & $-0.82314$\\
\enddata
 \end{deluxetable}
 
We use this analytic function to find the value of $\Gamma$ for each source suggested by its $h_{2}$ hardness ratio. Our function is defined as follows:
 \begin{equation}
     \Gamma = 1.2197e^{-1.6068h_{2}}-0.8004\, .
 \label{eq:gamlf}
 \end{equation}
The values of $\Gamma$ we obtain for our sources range from 0.66 to 4.46, with an average of 2.37 and a median of 2.04.
The individual values of $\Gamma$ used for each source can be seen in Table \ref{tab:rates}.
We then calculate the model flux in the 12--25~keV and 4--25~keV bands for each model spectrum, normalizing the models to ensure that the hard and full band count rates are consistent across models.
These fluxes and count rates are used to create a conversion factor between energy flux $F$ and count rate $r$
for values of $\Gamma = [0.5, 1.0, 1.5, 2.0, 2.5, 3.0, 3.5, 4.0]$ and use these points to create a second analytic function that relates $\Gamma$ to our conversion factor. 
We calculated these conversion factor relations for both the 4--25~keV and 12--25~keV count rates.

Although we use $\Gamma$ predicted by the $h_{2}$ hardness ratio in order to get luminosity estimates for our sources, we created and fit a function to relate $\Gamma$ to the $h_{1}$ ratio in order to determine how much the resulting luminosities would be affected.
We found that the $h_{1}$ ratio typically predicted lower/flatter values for $\Gamma$, which produced full band luminosities that differ by an average of $\sim20\%$ from those predicted by the $h_{2}$ ratio.
Most (66\%) of the $h_{1}$ predicted luminosities were larger than their $h_{2}$ counterparts.
These minor differences in luminosity did not affect the overall shape of the XLF, and are consistent with a turnover in the spectrum beyond 4--6~keV that the $h_{2}$ ratio would not capture.
We choose to use the values of $\Gamma$ predicted by the $h_{2}$ ratio based on its ability to describe the hard band luminosities for use in the hard band XLF.
We also choose to use this ratio because the spectra of black holes and neutron stars will differ most significantly above $\sim$10~keV there the $h_{2}$ ratio probes.
In this range, we would expect a hard state black hole to have $\Gamma = 1-2$ above $\sim$10~keV, while neutron stars would have a break that causes steepening to $\Gamma > 3$ in the same range.

We use the $h_{2}$-dependent $\Gamma$ values found using Equation \ref{eq:gamlf} to determine a 4--25~keV and 12--25~keV conversion factor for each source, and then multiply the count rate in a given band by its conversion factor to get an estimate of the flux in that band. 
The analytic functions for the conversion factors for the full and hard band fluxes both follow the form:
 \begin{equation}
     F_{i} = 10^{-11}~{\rm erg~cm}^{-2} \cdot r_{i} (Xe^{Y\Gamma}+Z) ,
 \label{eq:flux}
 \end{equation}
 where $F$ and $r$ are the flux (in erg s$^{-1}$ cm$^{-2}$) and count rate (in counts s$^{-1}$) respectively in energy band \textit{i}, and $X$, $Y$, and $Z$ are fitted parameters as defined in Table~\ref{tab:params}.

We use this flux to calculate the luminosity using a distance of 784~kpc as the luminosity distance \citep{Stanek98}.
These luminosities are representative of the intrinsic luminosity in the full and hard bands, due to the negligible effect of extinction at these energies.
The final count rates, full-band luminosities, and hardness ratios for each source can be found in Table~\ref{tab:rates}.

\section{Source Classification}
\label{sec:classification}

\begin{figure*}[ht!]
    \includegraphics[scale=0.43]{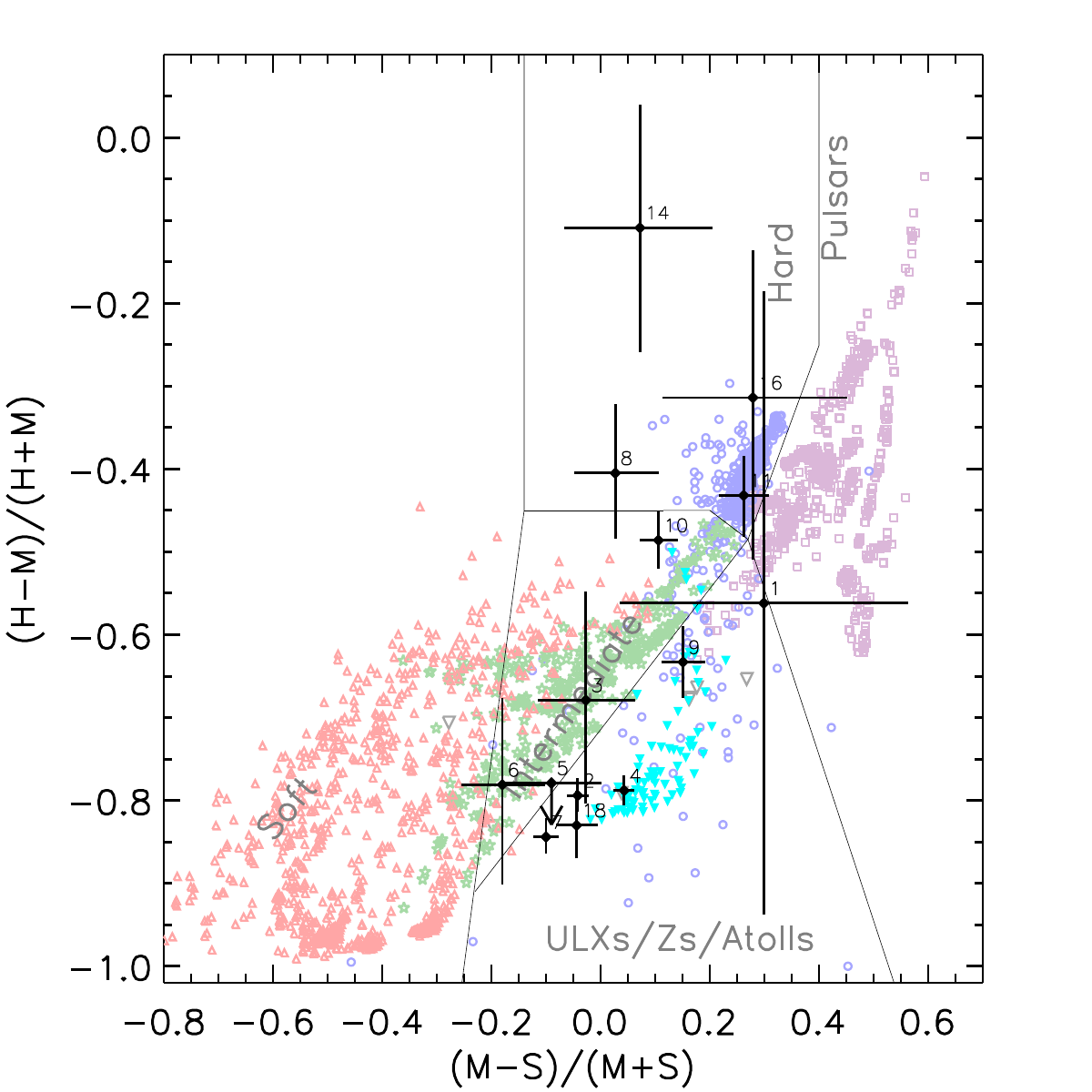}
    \includegraphics[scale=0.43]{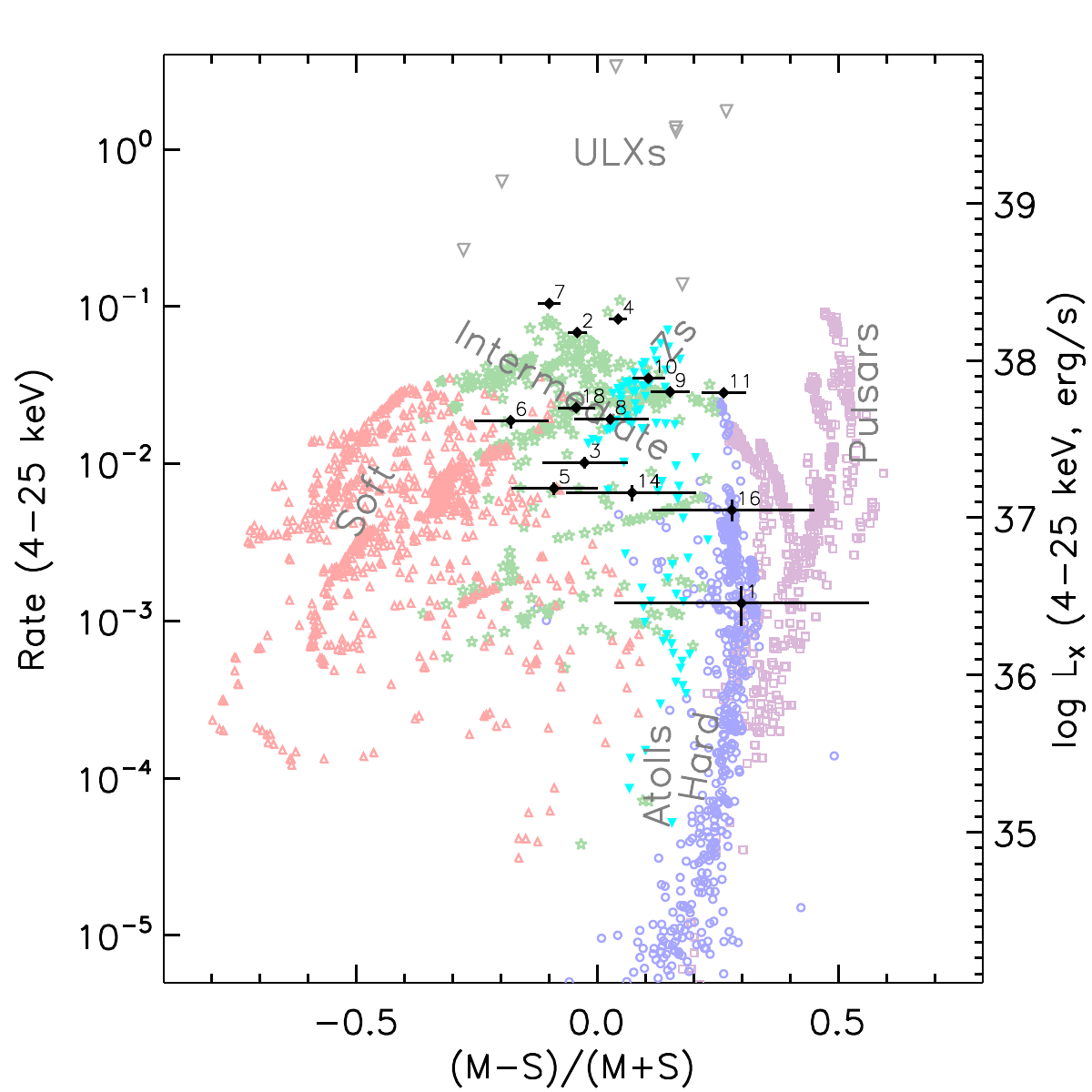}
    \caption{Diagnostic color-color (left) and color-intensity (right) diagrams for the XRB sources in our fields (black points).
    Error bars represent one sigma errors. 
    Background colorful symbols are simulated from \textit{RXTE} spectral fits of Galactic XRBs, and indicate different compact object types.
    These are: soft state black holes (red triangles), intermediate state black holes (green stars), hard state black holes (blue circles), accreting pulsars (purple squares), and Z and atoll type neutron stars (cyan triangles).
    Grey inverted triangles represent ULXs detected by \nustar.
    Classification regimes in the color-color diagram are separated by empirically-selected boundaries, shown as thin black lines. \label{fig:classplots}}
\end{figure*}

In order to classify our sources, we follow \cite{Wik14b, Yukita16, Lazzarini18, Vulic18, Lazzarini19} and \cite{Yang22} and create color-color and color-intensity diagrams using the hardness ratios and full band rates determined in Section \ref{sec:catalog}.
We impose boundaries on the color-color diagram to separate sources into one of five classifications, soft state black holes (BHs), intermediate state BHs, hard state BHs, non-magnetized neutron stars, and neutron star pulsars. 
These limits are determined empirically, and are based on the distribution of other classified sources.
The location of the BH-XRB and pulsar sources are determined by simulations of \textit{RXTE} spectral fits of Galactic XRBs (A. Zezas, private communication).
The Z/Atoll type NS points are based on spectral fits to galactic NS sources from \cite{Church14}.
The location of ULX points are based on \cite{Bachetti13, Rana15, Walton13} and \cite{Walton14}.
They can be seen in the background of the diagnostic color-color and color-intensity graphs in Figure \ref{fig:classplots}, and in \cite{Wik14b, Yukita16, Lazzarini18, Vulic18, Lazzarini19} and \cite{Yang22}.

Using the regions defined by these boundaries, we assume the 1$\sigma$ uncertainty ranges on the hardness ratios of each source represent a 2-D Gaussian probability distribution for the location of each source on the color-color diagram. 
This allows us to determine the rough probability that a given source would fall into each classification regime by simply integrating over the part of the distribution function within that source type regime.  
The classification probabilities reflect the statistical likelihood that a source overlaps with a given classification regime.
We choose to tentatively classify a source if at least 68\% of the Gaussian distribution overlaps with a single classification area.
In addition, we report the remaining probability that a source falls into any of the other classification regimes.

It should be noted that there are areas in both diagrams where sources with different accretion states overlap, including across different source regimes.
As a result, the classifications are considered tentative even when a source's location on the diagrams is very well-constrained.
Tentatively classified candidates are used in order to obtain the general distribution of BH and NS populations and to identify sources of interest for further analysis.
In future work, we hope to improve upon this classification method using a clustering algorithm, but this will require a larger number of confidently classified reference sources and is beyond the scope of the current work.

We also check for past classifications in other M31 X-ray surveys and for association with globular clusters.
We cross-correlate with the \cite{Stiele11}, \cite{Barnard14}, and \cite{Williams18} catalogs of X-ray and optical sources in M31 to identify past classifications, and check for association with globular clusters using the Revised Bologna Catalog \citep{Galleti04, Galleti09}.
Source 16 is associated with a star cluster in the Chandra-PHAT X-ray catalog \citep{Williams18}.
We consider a source coincident with a prior catalog source or globular cluster if the reported RA and Declination are within 10$^{\prime\prime}$ of our source position.

We find 9 sources that meet our 68\% requirement for tentative classification.
One of these, Source 8, is a known pulsar \citep{Esposito16,Zolotukhin2017}, and is thus not included in our count of new classifications.
The 9 sources that meet our classification limit are listed in Table \ref{tab:bestclass}, and the complete XRB source list can be found in Table \ref{tab:all}. 
Of the 14 XRB sources, 7 are associated with star clusters, and 13 are found in existing catalogs of X-ray sources in M31.

\section{Luminosity Functions}
\label{sec:luminosity}

We create X-ray luminosity functions (XLFs) in the full and hard bands, and include all sources detected above 2$\sigma$ from our data, as well as all disk sources detected above 2$\sigma$ from the Deep Paper.
We exclude all known AGN sources from both datasets.
In total, we add 40 disk sources from the Deep paper to the full band XLF, and 16 to the hard band XLF. 
Following the method described in Section \ref{sec:catalog}, all 14 of our XRB sources are detected at 2$\sigma$ in the full band. 
However, only 11 of the XRBs are detected at 2$\sigma$ above the background in the hard band. 
For those sources from the Deep Paper, we recalculate the luminosity of each source following the method described in Section \ref{sec:catalog} to ensure that all luminosities are calculated in the same way.
In total, we include 54 sources in the full band XLF, and 27 in the hard band XLF.

\movetabledown=45mm
\begin{rotatetable*}
\begin{deluxetable*}{ccc|CCCC|CCCC|c}
\tabletypesize{\scriptsize}
\tablecaption{\textit{NuSTAR} Rates and Colors for All Detected Sources \label{tab:rates}
}
\tablehead{
\colhead{} & \colhead{} & \colhead{} & \multicolumn{4}{c}{Count Rates} & \colhead{} & \multicolumn{2}{c}{{Hardness Ratios}} & \colhead{}\\ 
\colhead{}& \colhead{} &
\colhead{} & \colhead{S} &
\colhead{M} & \colhead{H} & \colhead{Full Band} & \colhead{L$_{\rm X}$\tablenotemark{a}} & \multicolumn{2}{c}{} & \colhead{} \\
\colhead{Source} & \colhead{RA (J2000.0)} &\colhead{Dec (J2000.0)} & \colhead{(4--6~keV)} & \colhead{(6--12~keV)} &\colhead{(12--25~keV)} & \colhead{(4--25~keV)} & \colhead{(4--25~keV)}&  \colhead{$\mathlarger{\frac{(M-S)}{(M+S)}}$} & \colhead{$\mathlarger{\frac{(H-M)}{(H+M)}}$} &\colhead{$\Gamma$} & \colhead{Type\tablenotemark{b}} \\
\colhead{ID} & \colhead{(deg)} & \colhead{(deg)} & \colhead{($10^{-3}$ cts s$^{-1}$)} & \colhead{($10^{-3}$ cts s$^{-1}$)} & \colhead{($10^{-3}$ cts s$^{-1}$)} & \colhead{($10^{-3}$ cts s$^{-1}$)} & \colhead{($10^{36}$ erg s$^{-1}$)} & \colhead{} & \colhead{} &\colhead{}}
\decimals 
\startdata
1 & 10.381818 & 41.295561 & \phn0.4\pm0.2 & \phn0.7\pm0.2 & 0.2\pm0.2 & \phn\phn1.3\pm0.4 & \phn\phn3.6\pm1.1 & +0.30\pm0.26 & -0.56\pm0.38 & 2.2 & XRB\\ 
2 & 10.429917 & 41.574700 & 33.8\pm1.0 & 31.1\pm0.9 & 3.5\pm0.4 & \phn68.4\pm1.4 & 147.4\pm3.0 & -0.04\pm0.02 & -0.79\pm0.02 & 3.5 & XRB \\ 
3 & 10.454713 & 41.019818 & \phn5.2\pm0.6 & \phn4.5\pm0.6 & 0.8_{-0.3}^{+0.4} & \phn10.2\pm0.9 & \phn24.7\pm2.2 & -0.03\pm0.09 & -0.68_{-0.12}^{+0.13} & 2.8 & XRB\\ 
4 & 10.564612 & 41.021803 & 37.9_{-1.4}^{+0.8} & 41.0\pm1.0 & 4.8\pm0.4 & \phn83.2_{-1.1}^{+2.0} & 179.3_{-2.4}^{+4.3} & +0.04\pm0.02 & -0.79\pm0.02 & 3.5 & XRB\\ 
5 & 10.590304 & 40.991983 & \phn3.6_{-0.4}^{+0.5} & \phn3.1\pm0.4 & 0.2\pm0.2 & \phn\phn7.0_{-0.6}^{+0.7} & \phn13.5_{-1.2}^{+1.4} & -0.09\pm0.09 & -0.91_{-0.00}^{+0.13} & 4.5 & XRB\\ 
6 & 10.643266 & 41.547174 & 10.6_{-1.3}^{+1.0} & \phn7.1\pm0.8 & 0.8_{-0.4}^{+0.5} & \phn18.8_{-2.0}^{+0.9} & \phn40.9_{-4.4}^{+2.0} & -0.18\pm0.08 & -0.78_{-0.12}^{+0.10} & 3.5 & XRB\\ 
7 & 10.716254 & 41.520568 & 55.4_{-1.9}^{+1.8} & 45.3_{-1.5}^{+1.6} & 3.3_{-0.4}^{+0.8} & 104.5_{-2.5}^{+2.6} & 214.6_{-5.1}^{+5.3} & -0.10\pm0.02 & -0.84\pm0.02 & 3.9 & XRB\\ 
8 & 10.756373 & 41.506220 & \phn7.7\pm0.9 & \phn8.1_{-0.8}^{+0.9} & 3.5_{-0.6}^{+0.7} & \phn19.2\pm1.4 & \phn63.6\pm4.6 & +0.03\pm0.08 & -0.40\pm0.08 & 1.5 & XRB\\ 
9 & 10.806662 & 41.122469 & 10.8_{-0.7}^{+0.8} & 14.6\pm0.8 & 3.2\pm0.4 & \phn28.6\pm1.1 & \phn73.2\pm2.8 & +0.15\pm0.04 & -0.63\pm0.04 & 2.5 & XRB\\ 
10 & 10.868304 & 41.131350 & 13.0_{-0.6}^{+0.8} & 16.2\pm0.7 & 5.5\pm0.4 & \phn34.9_{-1.1}^{+1.0} & 104.7_{-3.3}^{+3.0} & +0.11_{-0.03}^{+0.04} & -0.49_{-0.03}^{+0.04} & 1.9 & XRB\\ 
11 & 10.880112 & 41.179828 & \phn8.3\pm0.7 & 14.2\pm0.8 & 5.6\pm0.6 & \phn28.2_{-1.2}^{+1.1} & \phn90.4_{-3.8}^{+3.5} & +0.26_{-0.04}^{+0.05} & -0.43\pm0.05 & 1.6 & XRB\\ 
12 & 10.931000 & 41.479747 & \phn1.0\pm0.4 & \phn1.8_{-0.4}^{+0.5} & 0.9_{-0.4}^{+0.5} & \phn\phn3.5\pm0.8 & ... & +0.34_{-0.26}^{+0.23} & -0.32_{-0.25}^{+0.28} & 1.2 & PHAT galaxy \\ 
13 & 11.066224 & 41.517975 & \phn2.0\pm0.4 & \phn2.4\pm0.4 & 1.1_{-0.3}^{+0.4} & \phn\phn5.4\pm0.7 & ... & +0.11\pm0.14 & -0.40\pm0.16 & 1.5 & PHAT galaxy \\ 
14 & 11.173580 & 41.896168 & \phn2.0\pm0.4 & \phn2.3_{-0.4}^{+0.5} & 1.9_{-0.4}^{+0.5} & \phn\phn6.5\pm0.8 & \phn28.4\pm3.5 & +0.07_{-0.14}^{+0.13} & -0.11\pm0.15 & 0.7 & XRB \\ 
15 & 11.487267 & 41.809344 & \phn2.5\pm0.3 & \phn2.9_{-0.3}^{+0.4} & 1.3_{-0.2}^{+0.3} & \phn\phn6.7\pm0.5 & ... & +0.06\pm0.09 & -0.36\pm0.10 & 1.4 & PHAT galaxy \\ 
16 & 11.547964 & 42.141044 & \phn1.4\pm0.4 & \phn2.4_{-0.4}^{+0.5} & 1.4_{-0.4}^{+0.5} & \phn\phn5.1\pm0.8 & \phn18.5\pm2.9 & +0.28_{-0.16}^{+0.17} & -0.31_{-0.20}^{+0.18} & 1.2 & XRB\\ 
17a\tablenotemark{c} & 11.607031 & 42.073638 & 10.9\pm0.7 & \phn6.6\pm0.6 & 1.0\pm0.3 & \phn18.4\pm1.0 & ... & -0.25\pm0.05 & -0.74\pm0.08 & 3.2 & galaxy cluster\\
17b & 11.607031 & 42.073638 & \phn8.5\pm0.8 & \phn4.9\pm0.6 & 0.6\pm0.4 & \phn13.9\pm1.1 & ... & -0.26\pm0.07 & -0.78^{+0.13}_{-0.12} & 3.8 & galaxy cluster \\
18 & 11.609818 & 42.032944 & 11.4\pm0.6 & 10.4\pm0.6 & 1.0_{-0.2}^{+0.3} & \phn22.7_{-0.9}^{+0.8} & \phn47.0_{-1.9}^{+1.7} & -0.04\pm0.04 & -0.83\pm0.04 & 3.5 & XRB\\ 
19 & 11.699913 & 42.147866 & \phn0.9\pm0.3 & 1.1\pm0.3 & \phn0.3\pm0.3 & \phn\phn2.3\pm0.5 & ... & +0.06^{+0.23}_{-0.22} & -0.50^{+0.26}_{-0.28} & 1.9 & Stiele AGN\\
20 & 11.725742 & 42.173109 & \phn0.9\pm0.3 & \phn1.4\pm0.3 & 0.9_{-0.4}^{+0.2} & \phn\phn3.0\pm0.5 & ... & +0.26_{-0.19}^{+0.20} & -0.31\pm0.20 & 1.2 & PHAT galaxy\\
\enddata
\tablenotetext{a}{Luminosities are not reported for AGN/galaxy contaminants}
\tablenotetext{b}{XRB = candidate XRB source, PHAT galaxy = galaxy found in the Chandra-PHAT catalog \citep{Williams18}, Stiele AGN = marked as AGN in the Stiele catalog \citep{Stiele11}}
\tablenotetext{c}{17a and 17b are the same source. Source 17 appeared in two observations. We made independent measurements based on each observation, and thus report them separately as 17a and 17b.}
\end{deluxetable*}
\end{rotatetable*}

\begin{deluxetable*}{cc|ccccc|ccc}
 \tabletypesize{\scriptsize}
\tablecaption{{XRB Sources with Unique Classifications with $>$68\% Overlap} \label{tab:bestclass}}
\tablewidth{0pt}
\tablehead{ \colhead{} & \colhead{} & \multicolumn{5}{c}{Classification Overlap} & \multicolumn{3}{c}{Past Classifications}\\
\colhead{Source} & \colhead{Class\tablenotemark{a}} & \colhead{BH (soft)} & \colhead{BH (int.)} &\colhead{BH (hard)} & \colhead{NS (non-mag.)} & \colhead{NS (pulsar)} & \colhead{Catalog} & \colhead{Number} &\colhead{Class\tablenotemark{b}}}
\startdata
2 & NS (non-mag.) & 0.00 & 0.05 & 0.00 & \textbf{0.95} & 0.00 & Stiele & 694 & GlC \\
4 & NS (non-mag.) & 0.00 & 0.00 & 0.00 & \textbf{1.00} & 0.00 & Barnard & 109 & BHC \\
5 & NS (non-mag.) & 0.07 & 0.21 & 0.00 & \textbf{0.72} & 0.00 & Barnard & 146 & XRB \\
7 & NS (non-mag.) & 0.00 & 0.08 & 0.00 & \textbf{0.92} & 0.00 & Barnard & 327 & BHC \\
8 & BH (hard state)\tablenotemark{c} & 0.01 & 0.27 & \textbf{0.71} & 0.00 & 0.00 & Barnard & 378 & XRB \\
9 & NS (non-mag.) & 0.00 & 0.19 & 0.00 & \textbf{0.81} & 0.00 & Barnard & 427 & XRB \\
10 & BH (int. state) & 0.00 & \textbf{0.84} & 0.15 & 0.00 & 0.00 & Barnard & 471 & XRB \\
14 & BH (hard state) & 0.06 & 0.01 & \textbf{0.92} & 0.00 & 0.01 & ... & ... & ... \\
18 & NS (non-mag.) & 0.00 & 0.07 & 0.00 & \textbf{0.93} & 0.00 & Stiele & 1803 & GlC \\
\enddata
\tablenotetext{a}{NS (pulsar) = neutron star pulsar, NS (non-mag.) = Z or Atoll-type neutron star, BH = black hole in either the hard, soft, or intermediate (int) state}
\tablenotetext{b}{GlC = globular cluster, BHC = black hole candidate, XRB = X-ray binary, $<$hard$>$ = hard source with no other classification}
\tablenotetext{c}{Source 8 is a known pulsar \citep{Esposito16, Zolotukhin2017}. We have included our classification here for completeness, and explore the possible reasons for our misclassification in Section \ref{subsubsec:src8}.}
\end{deluxetable*}

We also create XLFs for sub-populations in the 4--25~keV band, based on whether or not sources were found to be associated with globular clusters, and whether they were classified as neutron stars or black holes.
In the case of the BH and NS XLFs, sources were separated by type based solely on their highest classification probability, even if it did not meet the 0.68 classification threshold chosen as the limit for tentative classification.
As a result, the fits for these sub-populations should be considered preliminary until there are enough high-probability classifications to produce their own fits.
Source 8 is included in the NS XLF, since it is a known pulsar.

In order to get completeness estimates for our XLFs, we calculate the faintest luminosity we could detect assuming a signal-to-noise ($S/N$) ratio of 2 at every location in our observations, as well as those from the Deep Paper. 
The $S/N$ is estimated using the number of net counts inside a $30^{\prime\prime}$ radius region.
For each band---12-25~keV and 4--25~keV---we calculate this luminosity using the median conversion factor between source counts and luminosity from our models in Section \ref{sec:catalog}.
The spread of possible conversion factors is such that the median values in each band are different from the lowest and highest values by less than a factor of 2.
For a given luminosity, the completeness is calculated by determining the fraction of the total solid angle where a source with that luminosity could be detected if present.

\begin{figure}
    \centering
    \includegraphics[scale=0.37]{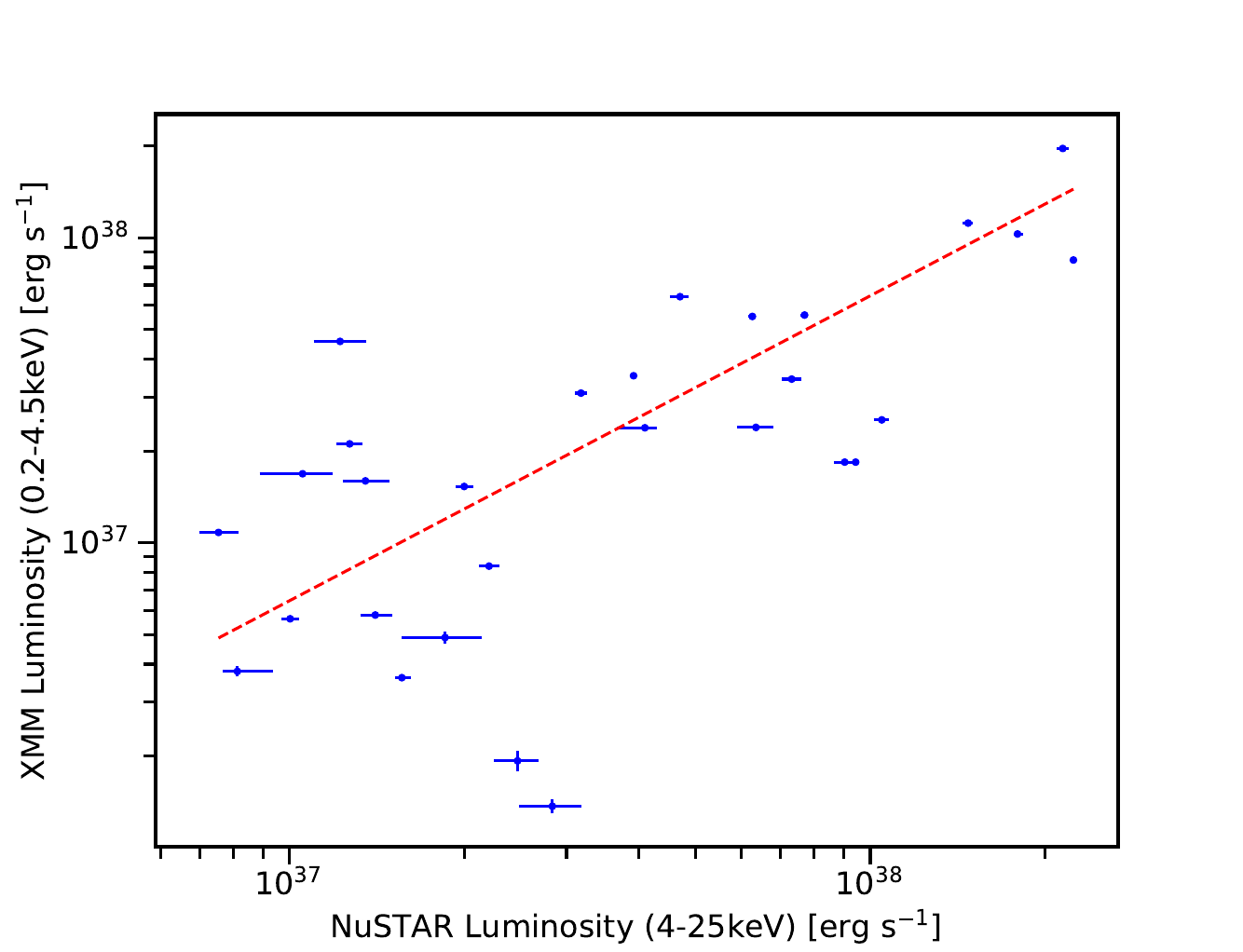}
    \caption{\nustar\ (4--25~keV) and \textit{XMM-Newton} (0.2--4.5~keV) luminosities for the 29 brightest disk sources in our data, along with their $1\sigma$ errors. 
    The expected scaling between \textit{XMM} and \textit{NuSTAR} luminosities, for a $\Gamma=1.75$ power law spectrum, is shown as the dashed red line.
    While there is appreciable scatter around this relation for individual sources (not surprising given the decade-long separation between these measurements), there is roughly symmetric scatter above and below it, suggesting a roughly stable XLF.
    }
    \label{fig:lumfactor}
\end{figure}

We check our completeness estimate by comparing our source number to the number of non-AGN sources detected with \textit{XMM Newton} in our regions by \cite{Stiele11}.
Based on our completeness estimates, we expect to detect 96\% of sources at or above L=7.5$\times$10$^{36}$ erg s$^{-1}$ in the 4--25~keV band.
For simplicity, we assumed little to no source variability, and matched the 29 sources in our data above this luminosity to sources in the \cite{Stiele11} data and used them to determine an average conversion factor between the 0.2-4.5~keV \textit{XMM} flux and our 4--25~keV \nustar\ flux.
Using WebPIMMS, we found that this conversion factor was consistent with the sources being well-represented by a power law with a slope of 1.75, which is similar to the slope of 1.7 assumed by \cite{Stiele11} when calculating their fluxes.
The unscaled \textit{XMM} and \nustar\ luminosities, along with the average conversion factor, can be seen in Figure \ref{fig:lumfactor}.

After converting the \cite{Stiele11} fluxes to the 4--25~keV band, 48 had scaled luminosities at or above our 96\% completeness limit of 7.5$\times$10$^{36}$ erg s$^{-1}$.
Of those 48, we found that two sources were background AGN and three were located too close to the edge of our \nustar\ observations to be detected, leaving 43 that we would reasonably expect to see.
Based on our 96\% completeness estimate, we would expect to see 41 of these 43 sources, but we detect only 29 sources above this limit.
However, eight of the 43 sources were detected in our sample at lower luminosities.
After accounting for these 8 sources, we are only missing four sources, which we consider to be indicative that our completeness estimate is reasonable.
This also suggests that the XLF is relatively stable.

We fit our data to two different differential luminosity functions, $dN/dL$: a broken power law and a single power law.
The broken power law is defined as:
\begin{equation}
    \frac{dN}{dL} = A\left\{
    \begin{array}{ll}
        L^{-\alpha_{1}}, & \text{for } L < L_{b} \\
        L^{-\alpha_{2}}L_{b}^{\alpha_{2}-\alpha_{1}}, & \text{for } L \geq L_{b}
    \end{array}
\right.\, ,
\label{eq:lfbpl}
\end{equation}
where A is the normalization, $L_{b}$ is the break luminosity, and $\alpha_{1}$ and $\alpha_{2}$ are the faint and bright indices respectively.
Similarly, the single power law is defined as:
\begin{equation}
    \frac{dN}{dL} = AL^{-\alpha_{1}}\, ,
    \label{eq:lfpl}
\end{equation}
where A is the normalization and $\alpha_{1}$ is the slope. 
$L$ and $L_{b}$ are in units of $10^{36}$ erg~s$^{-1}$.
During fitting, we correct the model XLF for completeness by:
\begin{equation}
    \left(\frac{dN}{dL}\right)_{i,corrected} = \left(\frac{dN}{dL}\right)_{i}\times c_{i}\, ,
\end{equation}
where $c_{i}$ is the completeness factor (ranging from 0 to 1) of the 
$i$th bin and $(dN/dL)_{i}$ is the value of the uncorrected model of the \textit{i}th bin given by Equation~\ref{eq:lfbpl} or \ref{eq:lfpl}.
This makes the corrected $(dN/dL)$ lower than the uncorrected one in order to match with the data, which we do not alter.

We construct the observed $dN/dL$ using 300 evenly spaced luminosity bins spanning the full observed luminosity range of the sample being fit, and compare this to our completeness-corrected models. 
Because most bins have very few or zero sources, we evaluated the likelihood using a modified C-statistic \citep{cash}:
\begin{equation}
    C = 2\sum_{i=1}^{n} M_{i} - N_{i} + N_{i}\log\left(\frac{N_{i}}{M_{i}}\right)\, ,
\end{equation}
where $M_{i}$ and $N_{i}$ are the model and observed number of sources, respectively, and summation occurs over the $n=300$ bins of luminosity.
In cases where the observed counts $N_{i}=0$, the $i$th term is equal to the model value $M_{i}$.

We find the best fit parameters and their uncertainties for each model using a Markov Chain Monte Carlo (MCMC) procedure, using the python \texttt{emcee} EnsembleSampler procedure \citep{emcee} to find the parameter set that maximizes the likelihood.
With this method, the fitting procedures are given an initial guess by the user, and a total of 100 walkers are generated by perturbing the initial guess according to a gaussian distribution with user-supplied standard deviations.

The value of the C-statistic is calculated for each parameter set for each walker, and we return the likelihood of each model using $\mathcal{L}$ = exp$\{-C/2\}$.
If the likelihood of the new trial from a walker, $\mathcal{L}_{new}$, is greater than that of the prior trial, $\mathcal{L}_{prior}$, we accept the new parameters and they are preserved as the prior trial for the next perturbation. 
Occasionally, new trials are accepted even in cases where $\mathcal{L}_{new}<\mathcal{L}_{prior}$.
In these cases, the move is accepted when the likelihood ratio $\mathcal{L}_{new}/\mathcal{L}_{prior}$ is equal to or larger than a randomly generated number between 0 and 1.
Otherwise, the prior trial is preserved for the next perturbation.
As a result, we get a number of accepted trials for each model equal to 100 times the total number of iterations we run (or, the number of iterations multiplied by the number of walkers).

We require that the indices for a given model be positive, and that the high-L slope be larger than the low-L slope in our broken power law fits.
We also require that the value of the break be below $10^{38}$~erg~s$^{-1}$, and also that it be no larger than the 3rd largest luminosity source.
This was done to avoid cases where the break was placed at or above the highest value of $L$ in the data.

We start the first run with an initial guess based on the appearance of the cumulative XLF, run it over 10,000 iterations, and choose large standard deviations for each parameter compared to the initial values in order to sample a large space.
We use this initial run as a burn-in, and take the median parameter values for each model and their standard deviations and use these as the initial guesses and widths for the next run, which we run over another ten thousand iterations.
Of those ten thousand further iterations, we thin the final sample by including every 50th accepted parameter set and exclude the first two hundred parameters as an additional burn-in.

We report the best fit model along with the median values of each parameter from the final sample set for each model, and report the 16\% and 84\% lower and upper confidence intervals.
We assess goodness of fit by comparing the C-statistic for the best fit model with the critical value as calculated according to \cite{Bonamente22}. 
The critical value is defined as:
\begin{equation}
    C_{crit} = E[C] + q\sqrt{Var[C]}
    \label{eq:Ccrit}
\end{equation}
where $E[C]$ and $Var[C]$ are calculated using Equation(s) 15.10 in \cite{Bonamente22}, and $q=0.5$, which corresponds to a probability $p$ of 0.68.
Models are considered acceptable when $C_{model}<C_{crit}$.
We show and discuss these XLFs in Sections \ref{subsec:xlf} and \ref{subsec:xlfcomparison}.

\section{Discussion}
\label{sec:discussion}

\subsection{Compact Object Classifications}
We are able to classify the compact object in 8 of our 14 XRB candidates with at least 68\% confidence, in addition to Source 8, which is a known pulsar \citep{Esposito16, Zolotukhin2017}. 
We report the hardness ratios, luminosities, and highest likelihood classifications of all of our XRB sources in Table~\ref{tab:all}, along with any matches with previous catalogs \citep{Stiele11, Barnard14}.
Three of these sources (4, 7, and 8) conflict with past classifications. 

\subsubsection{Source 4}
We classify Source 4 as a neutron star, but note that it was classified as a black hole candidate in the \cite{Barnard14} catalog.
Following \cite{Maccarone16}, we fit a variety of power law, multi-component blackbody disk (\texttt{diskbb}), and comptonization of soft photons in hot plasma (\texttt{comptt}) models to the spectrum in order to better constrain the nature of the object.

We find that the spectrum is best fit by a \texttt{comptt} model with $kT\sim2.1$~keV and $\tau=sim8.9$, which has a C-stat value of 531.98 for 556 degrees of freedom.
This model yields a 4--25~keV flux of $\sim2.0 \times 10^{-12}$ erg~s$^{-1}$~cm$^{-2}$, corresponding to a luminosity of $\sim1.5 \times 10^{38}$~erg~s$^{-1}$.
These values are all consistent with typical Z or atoll-type NS sources.

The spectrum can also be  fit by a \texttt{diskbb} model.
However, the normalization of the model predicts an unphysically small inner disk radius of only a few kilometers for a face-on disk. 
For a Schwarzschild BH, $R_{ISCO} = 90$km$(M/10M_{\odot})$.
Even for an extreme Kerr BH, $R_{ISCO}=15$km$(M/10M_{\odot})$ \citep{Remillard06}.
As a result, we find a multi-color disk model to be an unreasonable fit to the data.
Considering the fact that this source is also coincident with globular cluster Bol 82, we are confident in our classification of the source as a neutron star.

\subsubsection{Source 7}
We classify Source 7 as a neutron star, but it is classified as a black hole candidate in a globular cluster by \cite{Barnard14}.
We follow the same procedure to check our classification of Source 7 using its spectrum as we do for Source 4. 

Although the best fit models for Source 7 were \texttt{diskbb} models, the inner disk radius calculated based on the model normalization was unphysically small (typically 4~km), much like Source 4. 
None of the power law models we tried produced good fits to the data.
The \texttt{comptt} model produced a better fit than any of the power law models, but the value of $kT$ for the model was poorly constrained.

Following \cite{Church14}, we also fit a \texttt{bbody+cutoffpl} model with the absorption frozen to the Galactic value and a fixed power law index of $\Gamma=1.7$.
A good fit to the data was found, with a C-stat value of 496.34 for 493 degrees of freedom, and parameters similar to the best fit spectral models of high-L Z sources from \cite{Church14}.
The best fit parameters for the model can be found in Table \ref{tab:src7}.

\subsubsection{Source 8}
\label{subsubsec:src8}
Although we classified Source 8 as a hard state black hole, \cite{Esposito16} and \cite{Zolotukhin2017} have separately identified it as a pulsar with a relatively slow spin compared to other known globular cluster pulsars ($P_{\rm spin}=1.2$~s) using data from \textit{XMM-Newton}.
We investigate our classification in order to determine possible reasons why it may have been incorrectly classified by our methods.
\cite{Zolotukhin2017} use a \texttt{wabs(cflux*cutoffpl)} model with $\Gamma=0.2$, $E_{cut}=4.6$~keV, and $n_{H} = 3.79\times10^{20}$~cm$^{-2}$, yielding a reduced $\chi^2$ of 1.16 for 310 degrees of freedom. 
Using these same values and allowing only the normalization to vary, we obtain a similar fit statistic, with a reduced $\chi^2$ of 1.2 for 116 degrees of freedom.
We use the model predicted count rates to get estimates of the hardness ratios $h_{1}$ and $h_{2}$ used to create our color-color diagram.
Our fit for the model from \cite{Zolotukhin2017} predicts $h_{1} = 0.27\pm0.29$ and $h_{2}=-0.68\pm0.25$. 
This is consistent with our measured values of $0.03\pm0.08$ and $-0.40\pm0.08$ respectively, and the model-predicted ratios have error bars large enough to overlap with multiple classification areas of the color-color diagram.

We also calculate the 0.3--10~keV luminosity of the source using our best fit model from the \nustar\ data in order to compare it to the noted harder-brighter relationship reported by both \cite{Esposito16} and \cite{Zolotukhin2017}.
We find a model predicted 0.3--10~keV luminosity of $6.4\times10^{37}$ erg s$^{-1}$.
This is on the lower/softer side of the reported luminosities in \cite{Zolotukhin2017}, and may explain why the source falls into a softer area of parameter space than typical pulsar sources in our color-color diagnostics, causing us to classify it as a hard state black hole.
This possible overlap of different spectral states in certain areas of our diagnostic plots is the primary driver for our classifications being considered tentative, even in cases such as this where 100\% of the Gaussian distribution overlapped with a single classification regime.

\movetabledown=35mm
\begin{rotatetable*}
 \begin{deluxetable*}{c|CC|CC|c|cc|ccc}
\tabletypesize{\scriptsize}
\tablecaption{{Full XRB Source List} \label{tab:all}}
\tablehead{ \colhead{} & \multicolumn{2}{c}{Hardness Ratios} & \multicolumn{2}{c}{Luminosities [10$^{36}$erg s$^{-1}$]} & \colhead{} & \multicolumn{2}{c}{} & \multicolumn{3}{c}{Catalog Matches}\\
\colhead{ID} & \colhead{$\frac{M-S}{M+S}$} & \colhead{$\frac{H-M}{H+M}$} & \colhead{L$_{12-25~keV}$} &\colhead{L$_{4-25~keV}$} & \colhead{Globular Cluster \tablenotemark{a}} & \colhead{Class} & \colhead{Overlap\tablenotemark{b}} & \colhead{Catalog\tablenotemark{c}} & \colhead{ID} & \colhead{Class\tablenotemark{d}}}
\startdata
1 & +0.30\pm0.26 & -0.56\pm0.38 & \phn1.5\pm1.5 & \phn\phn3.6\pm1.1 & ... & NS (pulsar) & 0.39 & Barnard & 31 & [hard]\\ 
2 & -0.04\pm0.02 & -0.79\pm0.02 & 23.4\pm2.7 & 147.4\pm3.0 & Bol 45 & NS (non-mag.) & \textbf{0.95} & Stiele & 694 & GlC\\ 
3 & -0.03\pm0.09 & -0.68_{-0.12}^{+0.13} & \phn5.6_{-2.1}^{+2.8} & \phn24.7\pm2.2 & ... & NS (non-mag.) & 0.34 & Barnard & 50 & [hard]\\ 
4 & +0.04\pm0.02 & -0.79\pm0.02 & 32.1\pm2.7 & 179.3_{-2.4}^{+4.3} & Bol 82 & NS (non-mag.) & \textbf{1.00} & Barnard & 109 & BHC\\
5 & -0.09\pm0.09 & -0.91_{-0.00}^{+0.13} & \phn1.3\pm1.3 & \phn13.5_{-1.2}^{+1.3} & ... & NS (non-mag.) & \textbf{0.72} & Barnard & 146 & XRB\\
6 & -0.18\pm0.08 & -0.78_{-0.12}^{+0.10} & \phn5.34_{-2.7}^{+3.4} & \phn40.9_{-4.4}^{+2.0} & Bol 116 & NS (non-mag.) & 0.25 & Barnard & 198 & GlC\\ 
7 & -0.10\pm0.02 & -0.84\pm0.02 & 21.5_{-2.6}^{+5.2} & 214.5_{-5.1}^{+5.3} & Bol 135 & NS (non-mag.) & \textbf{0.92} & Barnard & 327 & GlC [BH]\\
8 & +0.03\pm0.08 & -0.40\pm0.08 & 27.1_{-4.6}^{+5.4} & \phn63.6\pm4.6 & Bol D91 & BH (hard state)\tablenotemark{e} & \textbf{0.71} & Barnard & 378 & GlC\\ 
9 & +0.15\pm0.04 & -0.63\pm0.04 & 23.0\pm2.9 & \phn73.2\pm2.8 & Bol 158 & NS (non-mag.) & \textbf{0.81} & Barnard & 427 & GlC\\
10 & +0.11_{-0.03}^{+0.04} & -0.49_{-0.03}^{+0.04} & 41.5\pm3.0 & 104.7_{-3.3}^{+3.0} & ... & BH (int state) & \textbf{0.84} & Barnard & 471 & XRB\\ 
11 & +0.26_{-0.04}^{+0.05} & -0.43\pm0.05 & 43.0\pm4.6 & \phn90.4_{-3.8}^{+3.5} & ... & NS (pulsar) & 0.22 & Barnard & 478 & [hard]\\ 
14 & +0.07_{-0.14}^{+0.13} & -0.11\pm0.15 & 15.6_{-3.3}^{+4.1} & \phn28.4\pm3.5 & ... & BH (hard state) & \textbf{0.92} & ... & ... & ... \\ 
16 & +0.28_{-0.16}^{+0.17} & -0.31_{-0.20}^{+0.18} & 11.1_{-3.2}^{+4.0} & \phn18.5\pm2.9 & ... & NS (pulsar) & 0.34 & Stiele & 1762 & [hard] (GlC)\tablenotemark{f}\\ 
18 & -0.04\pm0.04 & -0.83\pm0.04 & \phn6.5_{-1.3}^{+2.0} & \phn47.0_{-1.9}^{+1.7} & Bol 386 & NS (non-mag.) & \textbf{0.93} & Stiele & 1803 & GlC\\
\enddata 
\tablenotetext{a}{Globular cluster matches utilize the Revised Bologna Catalog \citep{Galleti04,Galleti09}for those matched with Bol clusters. Source 12 was reported to be incident with a globular cluster by \cite{Stiele11}.} 
\tablenotetext{b}{Bolded overlap indicates that the source is considered tentatively classified as described in Section \ref{sec:classification}.}
\tablenotetext{c}{Catalog name and ID from either the \cite{Barnard14} or \cite{Stiele11} catalogs.} 
\tablenotetext{d}{Square brackets indicate candidates/uncertain classification. hard = hard source with no other classification, GlC = globular cluster, BH = black hole, XRB = X-ray binary}
\tablenotetext{e}{ Source 8 known to be a pulsar \citep{Esposito16,Zolotukhin2017}}
\tablenotetext{f}{The cluster association for Source 16 was based on the PHAT catalog \citep{Williams18}}
\end{deluxetable*}
\end{rotatetable*}

\begin{deluxetable}{c|C|C|C|C|C}
\tabletypesize{\scriptsize}
\tablecaption{Source 7 \texttt{bbody+cutoffpl} parameters\tablenotemark{*} \label{tab:src7}}
\tablewidth{0pt}
\tablehead{\colhead{$N_{H}$} & \colhead{$kT_{bb}$ [keV]} &\colhead{$\Gamma$} & \colhead{$kT_{e}$ [keV]} & \colhead{$kT_{e}/kT_{bb}$} & \colhead{$kT_{e}/3kT_{bb}$}}
\startdata
0.17 & 1.45^{+0.28}_{-0.19} & 1.7 & 6.00^{+2.05}_{-2.11} & 4.14\pm 1.6 & 1.38 \pm 0.18
\enddata
\tablenotetext{*}{$kT_{bb}$ = blackbody temperature [keV], $\Gamma$ = index of cutoff power law, $kT_{e}$ = cutoff energy [keV]}
\end{deluxetable}

\subsection{Cumulative XLF Characteristics}
\label{subsec:xlf}
We are able to determine best fit models to differential XLFs (shown as cumulative XLFs) in both the full 4--25~keV and hard 12--25~keV bands following the procedure in Section~\ref{sec:luminosity}. 
In the 4--25~keV band, we also create and fit cumulative XLFs that separate between sources in globular clusters and in the field, as well as preliminary fits of NS-only and BH-only XLFs.

\begin{figure*}
\plottwo{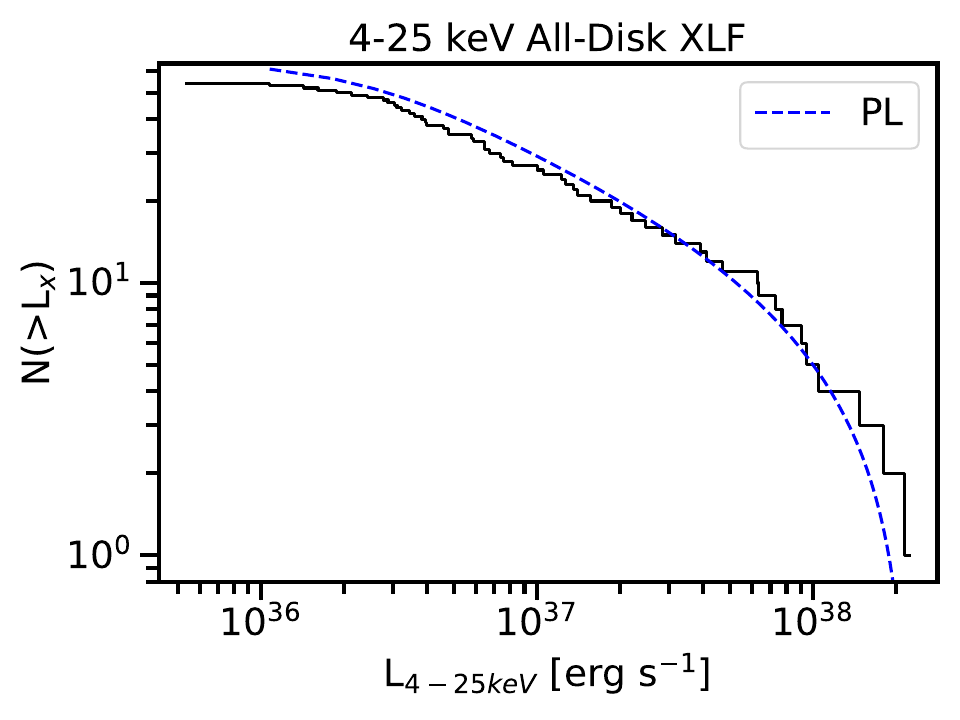}{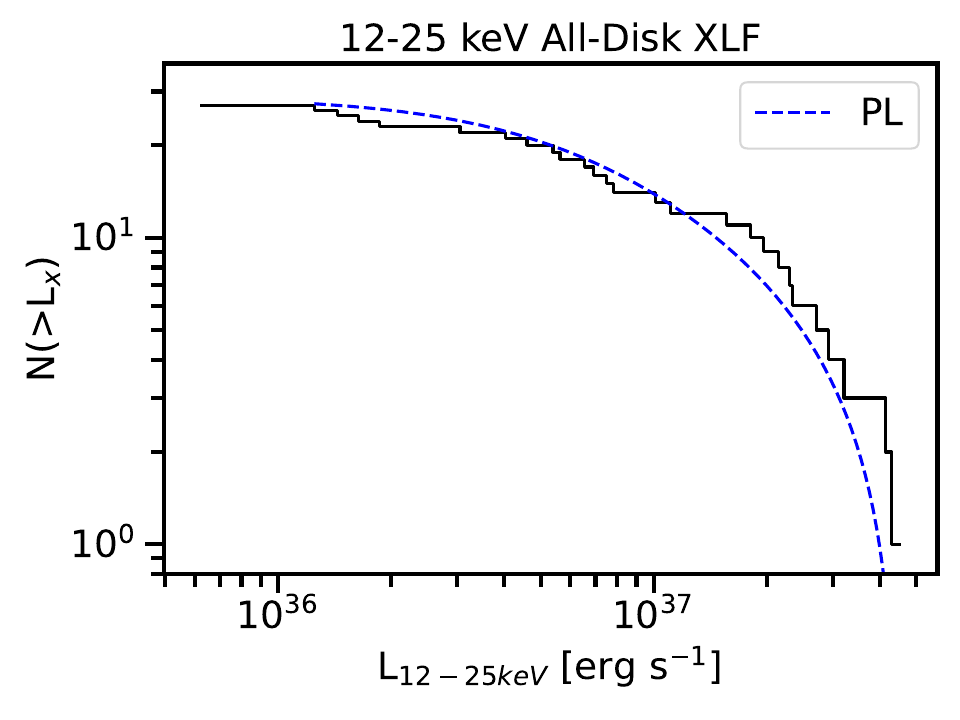}
    \caption{Cumulative XLFs for the M31 disk in the full (4--25~keV, left) and hard (12--25~keV, right) bands. The data is unbinned for plotting. Best fit power law models are plotted in blue. The data is not completeness-corrected, we instead make completeness corrections to the model during the fitting process, which is included in the models plotted here. \label{fig:xlfdisk}}
\end{figure*}

\begin{figure*}
\plottwo{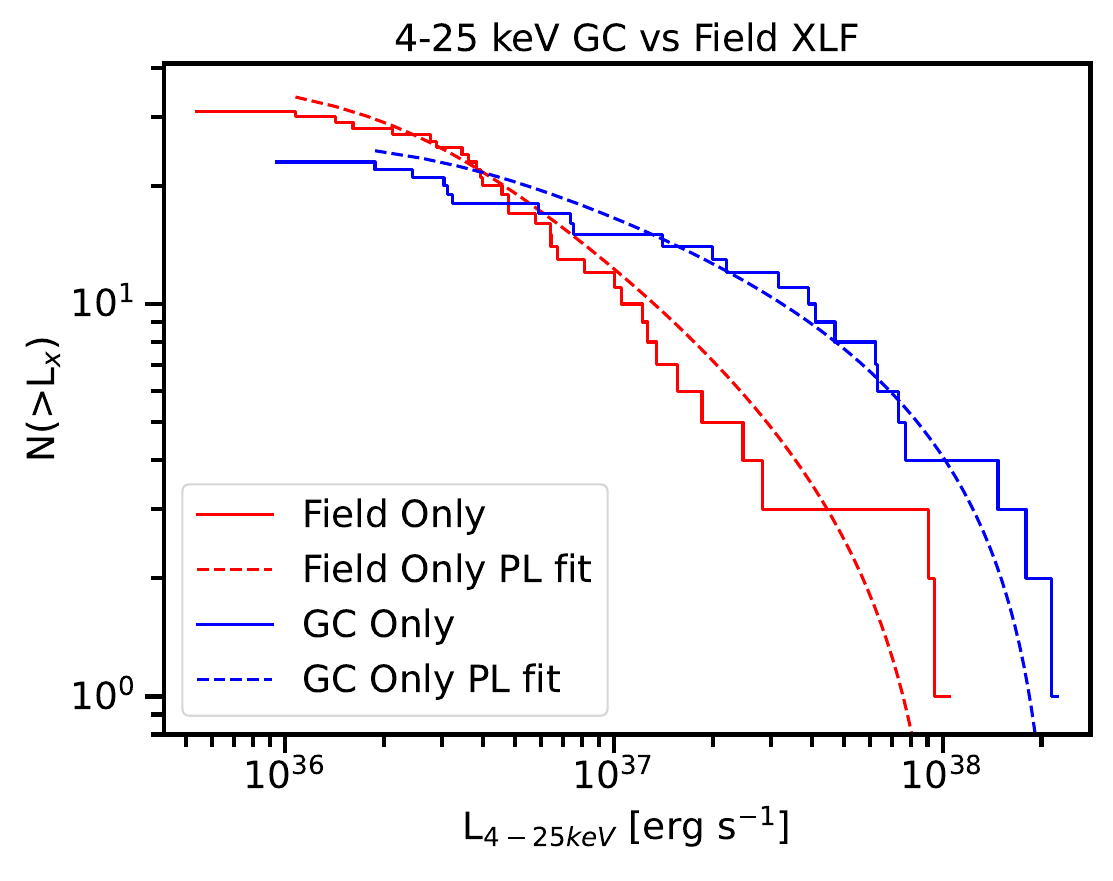}{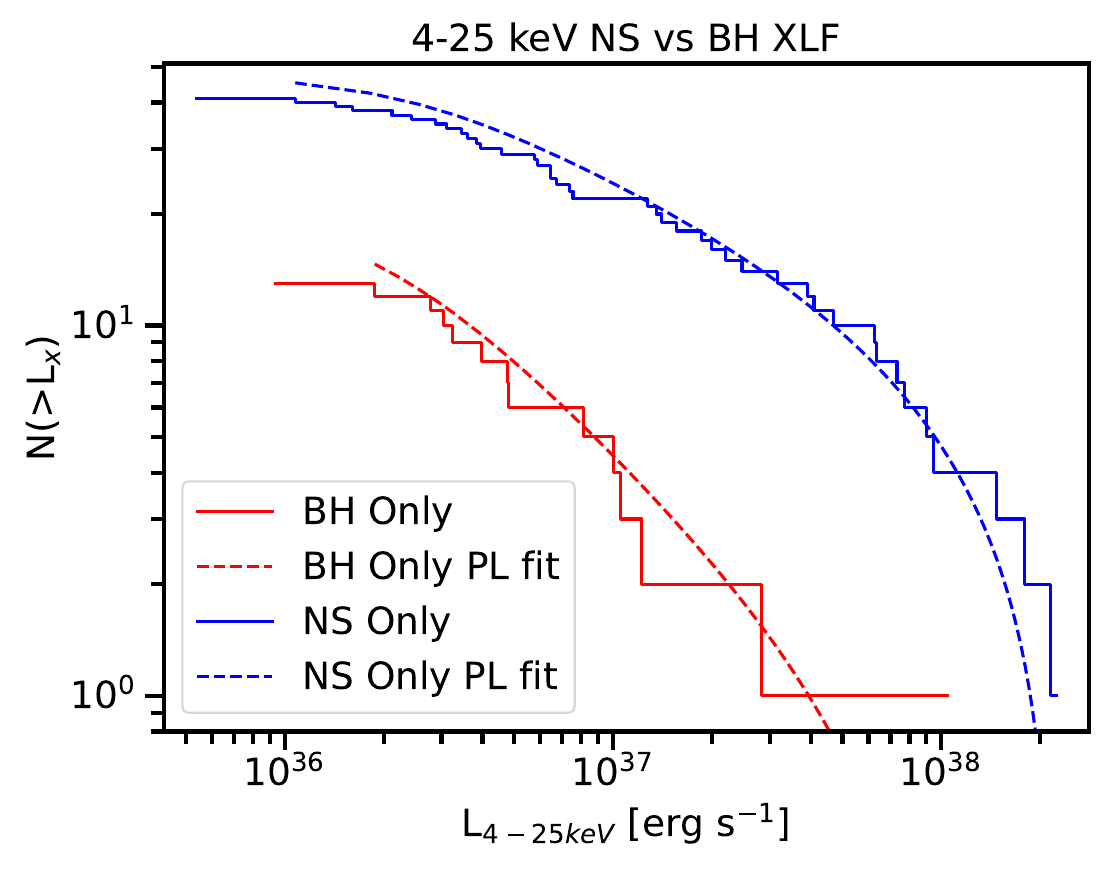}
    \caption{Cumulative XLFs and best-fit models for M31 disk sub-populations. The data are plotted with solid lines. Best fit models are plotted with dashed lines, and represent single power laws. We include sources in globular clusters (blue) and in the field (red) on the same plot to highlight the difference in shape. We do the same for the neutron star (blue) and black hole (red) sources. However, as noted in Section \ref{sec:classification}, the BH and NS classifications made with our diagnostic plots are tentative. In addition, we note that the black hole XLF includes only 5 sources with classification likelihoods above 68\%, and therefore is unlikely to be indicative of the population of black hole XRBs as a whole. As a result, the BH and NS XLFs should be considered preliminary. As in Figure \ref{fig:xlfdisk}, data is unbinned for plotting. \label{fig:XLFsubpops}}
\end{figure*}

In all of our XLF models, both the single and broken power laws produced acceptable fits to the data, with $C_{model}<C_{crit}$.
However, although the broken power laws produced better values of C-stat than single power laws, they did not produce a statistically significant improvement ($\Delta C<2$).
As a result, we report the best fit single power law parameters for each XLF, which can be found in Table~\ref{tab:XLFpars}.
The resulting completeness-corrected full and hard band cumulative X-ray luminosity functions (XLFs) and their best fit power law models can be found in Figures \ref{fig:xlfdisk} (all disk) and \ref{fig:XLFsubpops} (disk sub-populations).

\begin{deluxetable}{lCCC}[h!]
 \tabletypesize{\scriptsize}
\tablecaption{Best fit XLF model parameters \tablenotemark{a}\label{tab:XLFpars}}
\tablewidth{0pt}
\tablehead{\colhead{Subsample} & \colhead{$\alpha_{1}$} &\colhead{$C$} & \colhead{$C_{crit}$}}
\startdata
4-25~keV all disk &  1.32 \left(1.38^{+0.10}_{-0.11}\right) & 112.68 & 203.79 \\
12-25~keV all disk & 1.28 \left(1.49^{+0.22}_{-0.21}\right) & 118.59 & 141.05\\
4-25~keV GC Only & 1.03 \left(1.19\pm0.17\right)&  \phn99.20 & 129.37 \\
4-25~keV Field & 1.47 \left(1.59^{+0.17}_{-0.16}\right) & \phn85.11 & 152.04\\
4-25~keV NS Only &  1.23 \left(1.30\pm0.12\right) & 109.64 & 176.73 \\
4-25~keV BH Only\tablenotemark{b} & 1.74 \left(2.20^{+0.37}_{-0.33}\right) & \phn49.90 & 96.71 \\
\enddata
\tablenotetext{a}{Parameters are reported as Best Fit($50\%\pm34\%$), with the $50\%$ median and uncertainties calculated using MCMC chains marginalized across all parameters. $C$ and $C_{crit}$ are the C-stat value of the best fit and the critical C-stat value respectively. $C_{crit}$ is calculated according to \cite{Bonamente22}. Fits are considered acceptable when $C<C_{crit}$.} 
\tablenotetext{b}{The BH-only XLF is comprised of only 13 sources, and is therefore unlikely to be indicative of the population of BH XRBs as a whole.}
 \end{deluxetable}

\subsection{Comparison to the Bulge XLF}
\label{subsec:Bulgexlf}
The Deep Paper made characterizations of the XLF of the bulge of M31 in the 4--25 and 12--25~keV bands, both of which were best fit by broken power laws. 
These XLFs for the bulge, as well as their best fit parameters, are included for reference in Appendix \ref{sec:bulge}.
The breaks in the bulge fits are consistent with breaks found in prior characterizations of the cumulative XLF in the Milky Way \citep{Voss10, Revnivtsev11, Doroshenko14} and M31 \citep{Primini93, Kaaret02, Kong02, Greening09, Vulic16}.
\cite{Revnivtsev11} suggest that for LMXBs, a break at $\sim2\times10^{37}$ can be explained by a change in the donor star in the binary, with evolved secondaries (giants) above the break, and main sequence stars below the break.
This could explain why we see a break in the bulge XLF, but not in the disk.
If the break is primarily caused by LMXBs, it would be less likely to appear in the disk XLF, which should include a lower fraction of LMXBs than the bulge due to the respective ages of the populations.

However, \cite{Voss10}, \cite{Lutovinov13}, and \cite{Doroshenko14} find that breaks are also present in HMXB XLFs in the Milky Way. 
The \cite{Voss10} model in particular finds a break at the same luminosity as the \cite{Revnivtsev08} model, although the errors on the break measurement are quite large. 
This could indicate that breaks in the XLF are due instead to factors affecting both LMXB and HMXB populations.

Only 5 of the 54 sources included in our full band XLF are listed as HMXBs in the Chandra-PHAT catalog\citep{Williams18}.
As a result, we are not yet able to explore whether a difference in XLF shape results from differences in the relative HMXB and LMXB populations. 
New data to be taken in \nustar\ cycle 9 will have significant overlap with the Panchromatic Hubble Andromeda South Treasury \citep[PHAST:][]{PHAST}, which may yield more confidently classified HMXB sources in the disk.

 \begin{deluxetable*}{lccc|ccc}[ht!]
 \tabletypesize{\scriptsize}
\tablecaption{Comparison to Milky Way  XLFs \label{tab:compxlfs}}
\tablewidth{0pt}
\tablehead{\colhead{} & \multicolumn{3}{c}{LMXB} & \multicolumn{3}{c}{HMXB} \\
\colhead{Survey} & \colhead{$\alpha_{1}$} &\colhead{$\alpha_{2}$} &\colhead{L$_{b}$[10$^{36}$erg s$^{-1}$]} & \colhead{$\alpha_{1}$} &\colhead{$\alpha_{2}$} &\colhead{L$_{b}$[10$^{36}$erg s$^{-1}$]}}
\startdata
Doroshenko+ 2014 &  $0.9^{+0.2}_{-0.4}$ & $2.6^{+3.0}_{-0.9}$ & $8^{+7}_{-6.5}$ & $0.3^{+0.8}_{-0.2}$ & $2.1^{+3}_{-0.6}$ & $0.55^{+4.6}_{-0.28}$ \\
Voss+ 2010 & $0.9^{+0.4}_{-0.3}$ & $2.4^{+0.4}_{-0.7}$ & $3.0^{+1.8}_{-1.6}$ & $1.3^{+0.2}_{-0.2}$ & $>2$ &  $25^{+200}_{-23}$ \\
Lutovinov+ 2013 & ... & ... & ... &  $1.4\pm0.19$ & $>2.2$ & $2.5^{+3.7}_{-2.3}$\\
Grimm+ 2002 & $1.26\pm0.08$ & ... & ... & $1.64\pm0.15$ & ... & ... \\
Revnivtsev+ 2008 & $1.13\pm0.13$ & ... & ... & ... & ... & ... \\
\enddata
 \end{deluxetable*}

\subsection{Comparison to Milky Way XLFs}
\label{subsec:xlfcomparison}

There are a number of existing XLF models for LMXBs and HMXBs in the Milky Way, which we use to compare with our best fit M31 XLFs.
The best fit parameters for these models are summarized in Table \ref{tab:compxlfs}.
Though we expect the M31 disk population to be dominated by HMXBs, we compare it with both HMXB and LMXB XLFs in the Milky Way.

When comparing our single power laws to the broken power laws found by \cite{Voss10}, \cite{Lutovinov13}, and \cite{Doroshenko14}, we compare with the slope that overlaps the most with the range of luminosities in our fits.

In the case of the LMXB fit from \cite{Doroshenko14}, the break is located at $8^{+7}_{-6.5}\times10^{36}$ erg s$^{-1}$, which lies roughly in the center of our luminosity range.
In this case, we compare our best fit values with both slopes.
When determining the uncertainty range for our model parameters, we consider the smallest range that contains the best fit values along with the median and $16-84\%$ values from MCMC. 
We consider our models comparable to existing models when the best fit parameters are consistent within errors.

Based on this criteria, we find that our all-disk full band XLF is consistent with the low-L slope of the \cite{Voss10} HMXB XLF, where we have the greatest overlap in luminosity.
It is also consistent with the \cite{Grimm02} LMXB XLF, and only slightly steeper than the HMXB XLF.
Our all-disk hard band XLF is consistent with the high-L slopes of both the \cite{Doroshenko14} HMXB and LMXB XLFs, both of the low-L slopes of the \cite{Voss10} XLFs, and both of the \cite{Grimm02} XLFs. 
Both the hard and full band XLFs are slightly steeper than the \cite{Revnivtsev08} model.

We compare our 4--25~keV globular cluster XLF fit to the LMXB XLF models for the Milky Way, because the stellar companions in globular clusters are most likely low mass stars.
Our fit is consistent with the \cite{Grimm02} and \cite{Revnivtsev08} models, although the \cite{Revnivtsev08} XLF contains no sources above 10$^{37}$ erg s$^{-1}$.
The \cite{Voss10} LMXB XLF is fit by a broken power law, but we have only two sources below the break value, so we compare our model with their high-L index. 
We find a shallower index for the GC XLF than the \cite{Voss10} LMXB XLF, although their XLF contains very few sources above 10$^{37}$ erg s$^{-1}$.

\subsection{XLF Sub-Populations}

The cumulative XLFs of sub-populations in the disk are shown in Figure \ref{fig:XLFsubpops}, and the best fit parameters are listed in Table \ref{tab:XLFpars}.

We find that the XLF of sources in globular clusters is shallower than the field XLF.
This is consistent with \cite{Lin15}, who constructed XLFs of LMXBs in NGC 3115 using \textit{Chandra} data, and observed that sources in globular clusters produced a flatter XLF than field sources.  
The best fit XLF models for our neutron star and globular cluster sub-populations are also consistent with each other. 
While there are more neutron star sources than globular cluster sources (41 vs 23 respectively), 15 of the 23 globular cluster sources are  classified as neutron stars with probabilities above 0.68, and only 2 are classified as black holes.  
This is in agreement with the findings in \cite{Maccarone16}, which suggested that bright XRBs in globular clusters in M31 are most likely neutron stars.

Although we create an XLF for black hole sources in the disk, the low number of sources included (13) does not allow us to make any confident assertions about possible differences in shape between BH and NS XLFs. 
Though the preliminary fit suggests that the black hole XLF is steeper, only 5 of the sources in the XLF are classified as black holes with Gaussian overlap above 0.68.
Due to the tentative nature of these classifications, the BH and NS XLFs are considered preliminary.
Additional data of the southern disk to be taken in \nustar\ Cycle 9 may increase the number of candidate black holes in the disk, possibly allowing for more robust modelling of the 4--25~keV black hole disk XLF.

\subsection{XLF vs SFR and Stellar Mass}
\label{subsec:sfr}
We use our best fit 4--25~keV power law XLF model to obtain an estimate of the integrated luminosity $L_{X}$ and compare with the predicted $L_{X}$ from scaling relationships in \cite{Vulic18} and \cite{Lehmer19}.
In these models, the total integrated luminosity is a combination of HMXB and LMXB contributions.
These individual contributions are based on the following scaling parameters:
\begin{equation}
    \alpha_{LMXB} \equiv \frac{1}{M_{*}}\int_{L_{0}}^{L_{max}} \frac{dN}{dL}dL = L_{X}(LMXB)/M_{*} 
\end{equation}
\begin{equation}
    \beta_{HMXB} \equiv \frac{1}{SFR}\int_{L_{0}}^{L_{max}} \frac{dN}{dL}dL = L_{X}(HMXB)/SFR\, ,
\end{equation}
where $M_{*}$ is the stellar mass of the galaxy and SFR is the total galaxy star formation rate.
\cite{Vulic18} find $\alpha = 3.56\pm1.163 \times 10^{28}$ erg s$^{-1}$ M$_{\odot}^{-1}$ and $\beta = 1.902 \pm 0.837 \times 10^{39}$ erg s$^{-1}$ (M$_{\odot}$ yr$^{-1}$)$^{-1}$ using data from 8 Milky Way type galaxies with \nustar\ in the 4--25~keV range.
\cite{Lehmer19} find log$_{10}\alpha = 29.15^{+0.07}_{-0.05}$ and log$_{10}\beta = 39.73^{+0.15}_{-0.10}$ using \textit{Chandra} observations of 38 galaxies in the 0.5--8~keV range.
Using these scaling relationships, we find a total predicted $L_{X} = \alpha M_{*} + \beta SFR$ and compare with the total $L_{X}$ we calculate using our XLF model.

We make use of the global average SFR for the disk over the past 100 Myr from \cite{Lewis15}, which defines the disk according to the D25 ellipse from \cite{Paz07}.
We scale this again based on the areal coverage of our \nustar\ fields compared to that D25 ellipse, and obtain a star formation rate of 0.22 $M_{\odot}$~yr$^{-1}$.
Similarly, we use the stellar mass reported by \cite{Williams17} to estimate the stellar mass within our fields, yielding a value $M_{*} = 3.33 \times 10^{10}M_{\odot}$.

In order to calculate $L_{X}$ from our model, we use our best fit power law model to calculate:
\begin{equation}
    L_{X} \equiv \int_{L_{0}}^{L_{max}} \frac{dN}{dL}L dL\, ,
\end{equation}
where $L_{0} = 1\times10^{36}$~erg~s$^{-1}$ and $L_{max} = 5\times10^{40}$~erg~s$^{-1}$ to match \cite{Lehmer19}.
We calculate $L_{X}$ for the power law parameters from the $68\%$ highest likelihood fits from the MCMC chains described in Section \ref{sec:luminosity}, and use this range to determine our $16\%-84\%$ uncertainty on $L_{X}$.
Because our XLF covers a different range of energy than the one used to obtain the scaling relationship from \cite{Lehmer19}, we report scaled and unscaled values of $L_{X}$ in Table~\ref{tab:intlx} for better comparison with the prediction based on the 0.5--8~keV relationship.

We use \texttt{XSpec} to determine the ratio of fluxes in the 0.5--8~keV and 4--25~keV bands using a dummy response and the median spectral model parameters used by \cite{Lehmer19}.
For any galaxy where individual spectral fits were not possible, they use a \texttt{tbabs$_{Gal}\times$tbabs$_{int}\times$pow} model, with an intrinsic absorption component $\log_{10}N_{H,{\rm int}}=21.3$, the Galactic column density $N_{\rm H,Gal}$ as the other absorption component, and a power law index $\Gamma=1.7$. 
Using these parameters and the column density toward M31 ($N_{\rm H,Gal}=1.69\times10^{21} \text{cm}^{-2}$) \citep{HI4PI}, we find that the 4--25~keV luminosity is only 1.38 times higher than the 0.5--8~keV luminosity.
We adjust the 4--25~keV $L_{X}$ we obtain from our best fit down by this factor in order to obtain the scaled $L_{X}$.
We find values of $40.75^{+2.14}_{-1.24}$ and $40.84^{+2.25}_{-1.30}$ for the scaled and unscaled model $L_{X}$ respectively.

We also calculate a total $L_{X}$ based only on the data by summing up each of the 54 individual source luminosities, including disk sources from the Deep Paper. 
This yields log$_{10}L_{X}$=39.26, which is in agreement with the \cite{Vulic18} prediction, but slightly lower than the value predicted by the \cite{Lehmer19} parameters.
The predicted log$_{10}L_{X}$ based on the \cite{Vulic18} parameters are $39.21\pm0.24$, and the \cite{Lehmer19} scaling relationship predicts $39.77^{+0.17}_{-0.11}$.

The scaled and unscaled $L_{X}$ based on our integrated X-ray luminosity models are slightly higher than both the \cite{Vulic18} and \cite{Lehmer19} values. However, we are consistent within errors with the values predicted by the scaling relationships in \cite{Lehmer19}. 

\begin{deluxetable}{c|c|C}
 \tabletypesize{\scriptsize}
\tablecaption{Integrated X-ray Luminosity \label{tab:intlx}}
\tablewidth{0pt}
\tablehead{\colhead{Calculation Method} & \colhead{Energy Range [keV]} &\colhead{log$_{10}L_{X}$} }
\startdata
Prediction from \cite{Vulic18} & 4--25 & 39.21\pm0.24 \\
Prediction from \citep{Lehmer19} & 0.5--8 & 39.77^{+0.17}_{-0.11} \\
Model-derived scaled $L_{X}$ & scaled\tablenotemark{a} 4--25 & 40.75^{+2.14}_{-1.24}  \\
Model-derived unscaled $L_{X}$ & 4--25 & 40.84^{+2.25}_{-1.30}  \\
Summed data\tablenotemark{b} & 4--25 & 39.26
\enddata
\tablenotetext{a}{The scaled $L_{X}$ is adjusted as described in Section \ref{subsec:sfr}.}
\tablenotetext{b}{In addition to getting a total $L_{X}$ based on our best fit model, we also sum up the luminosities of each of our sources}
\end{deluxetable}

\section{Summary}

We used 10 $\sim$40~ks \nustar\ observations in the disk of M31 to study the XRB population and determine their compact object types and accretion states.
We tentatively classify 8 of 14 XRBs in the M31 disk, with a distribution of 2 black holes and 6 neutron stars.
We create and characterize the 4--25~keV cumulative XLF, and present the first characterization of a hard (12--25~keV) band cumulative  XLF for M31.
We found that the full band XLF was best fit by a power law with index $\alpha_{1} =1.32 \left(1.38^{+0.10}_{-0.11}\right)$.
The 12--25~keV XLF was best fit with a power law with index $\alpha_{1} =1.28 \left(1.49^{+0.22}_{-0.21}\right)$.
In addition, we separately characterize the XLFs of globular and field sources in these bands and find that the GC XLF is flatter than the field XLF (power law indices of $\sim$1.03 and $\sim$1.47, respectively). 
We make characterizations of the shape of a NS-only XLF  best fit by a power law with $\alpha_{1}=1.23 \left(1.30\pm0.12\right)$, and produce a preliminary BH-only XLF best fit by a power law with index $\alpha_{1} =1.74 \left(2.20^{+0.37}_{-0.33}\right)$. 
These XLFs are created by combining the deep and shallow coverage of the M31 disk. 
With additional data, it may be possible to determine whether or not there is a difference in XLF shape based on compact object type.
The consistency between the globular cluster and NS XLFs may suggest that bright XRBs in globular clusters in M31 are most likely NS sources.

If further data do indicate a difference in XLF shape by compact object, it may be possible to break down galaxy XLFs into contributions from BH and NS populations, allowing for broad population characterization in galaxies where source confusion or a lack of individually resolved sources prevents individual compact object classification.

Furthermore, the shape of HMXB XLFs and the total luminosity of HMXB populations in galaxies are linked to parameters like galaxy star formation rate, stellar mass, and metallicity \citep{Kilgard02, Gilfanov04,Mineo12,Vulic18,Lehmer19,Lehmer21,Lehmer2020}, which are important in models of accreting binary evolution \citep{Fragos09,Tzanavaris13, Zuo14,Giacobbo18,Misra23}. 
Further characterization of X-ray point source populations in galaxies allows for better comparisons to the outputs of these models. 
As a result, XRB population studies are important indicators of the population of gravitational wave sources.  
In the future, X-ray probes with high-E (E$>$10~keV) sensitivity such as HEX-P \citep{Madsen19, HEXP} will allow for this type of characterization using energies above 10~keV with even better spatial resolution, expanding our reach to more distant galaxies where \nustar\ is unable to resolve individual sources.

Future work will add classifications of XRBs in the southern disk using upcoming \nustar\ coverage, with the additional goals of identifying a statistically significant break in the full and hard band XLFs and creating distinct BH and NS XLFs.

\section*{Acknowledgments}
\begin{acknowledgments}
We thank the referee for helpful comments that improved the paper.
This work was made possible with funding from the NuSTAR Guest Observer program under grant 80NSSC17K0617 and JPL subcontract RSA1626360.
This work made use of data from the NuSTAR mission, which is led by the California Institute of Technology, managed by the Jet Propulsion Laboratory, and funded by the National Aeronautics and Space Administration.
Additionally, this research has made use of data and software provided by the High Energy Astrophysics Science Archive Research Center (HEASARC), which is a service of the Astrophysics Science Division at NASA/GSFC and the High Energy Astrophysics Division of the Smithsonian Astrophysical Observatory.
This work also made use of Astropy---a community-developed core Python package for Astronomy (Astropy Collaboration, 2013)---, numpy---Van der Walt and Colbert 2011, Computing in Science \& Engineering 13, 22---, and scipy---a community-developed open source software for scientific computing in Python \citep{astropy:2013,astropy:2018,astropy:2022,SciPy,numpy:2011,numpy}.
Finally, we would like to thank Dr. Ram\'{o}n Barthelemy for suggested edits, and Dr. Massimillano "Max" Bonamente for providing clarification on the calculation of our expected and critical values of the C-statistic. 
\end{acknowledgments}

\vspace{5mm}
\facilities{\textit{NuSTAR}}

\software{astropy\citep{astropy:2013,astropy:2018,astropy:2022}, emcee \citep{emcee}, numpy \citep{numpy}, scipy \citep{SciPy}}

\appendix
\restartappendixnumbering
\section{Bulge Fits}
\label{sec:bulge}

\begin{figure*}[h!] \plottwo{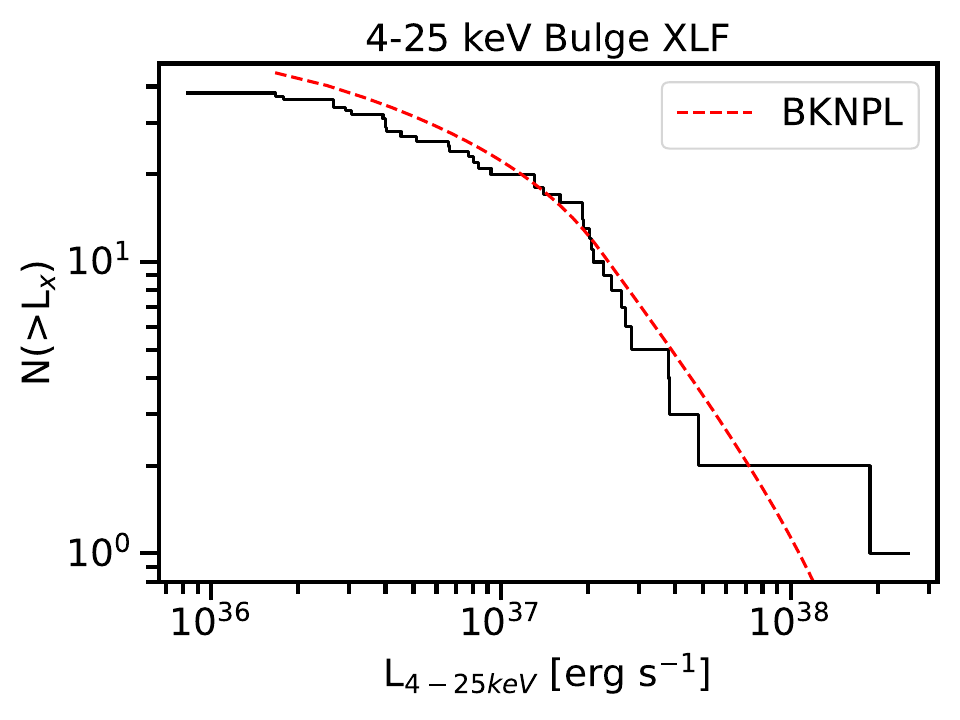}{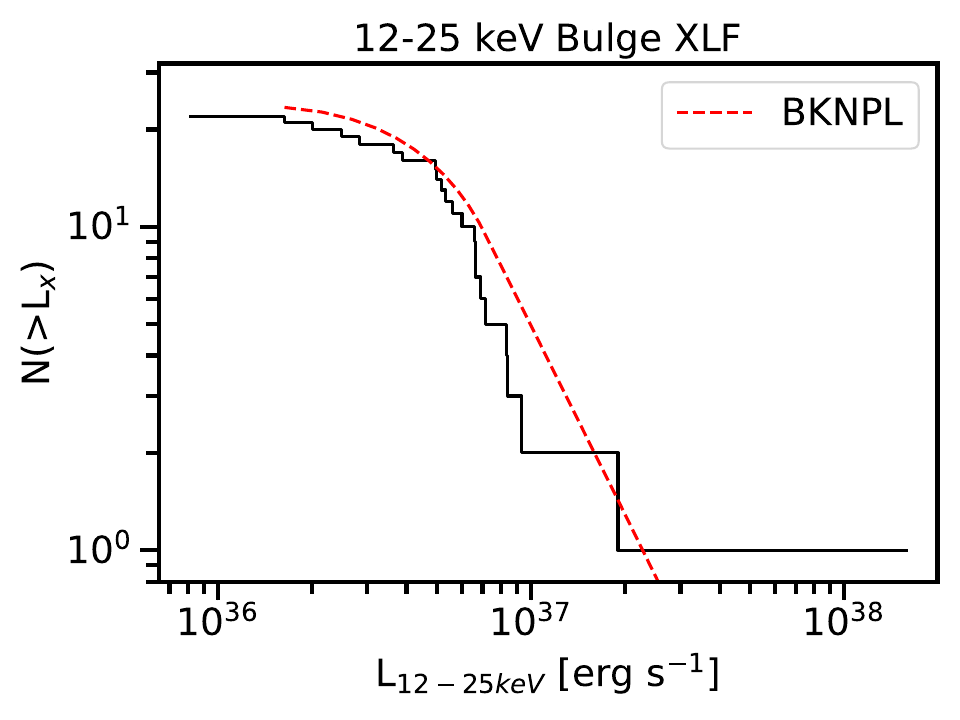}
\caption{The cumulative full band (left) and hard band (right) XLFs of the bulge and their best fit models.
The data are shown by solid black lines, and the best fit models as dashed red lines. The data is not completeness corrected. Instead, we correct the models for completeness as described in Section~\ref{sec:luminosity}. Best fit parameters are listed in Table \ref{tab:bulgexlfs}. \label{fig:bulgexlf}}
\end{figure*}

The bulge XLFs from the Deep Paper are reproduced here for reference.
The best fit models were determined according to the same procedure as the one described in Section~\ref{sec:luminosity}.
The full and hard band bulge XLFs are both best fit by broken power law models, which are shown in Figure~\ref{fig:bulgexlf}. The best fit parameters for the models are listed in Table~\ref{tab:bulgepars} and are discussed in brief in Section~\ref{subsec:Bulgexlf}.

\begin{deluxetable*}{lCCCCC}[h!]
    \label{tab:bulgexlfs}
    \tabletypesize{\scriptsize}
    \tablecaption{Bulge XLF Parameters\tablenotemark{a}\label{tab:bulgepars}}
\tablewidth{0pt}
\tablehead{\colhead{Energy Band} & \colhead{$\alpha_{1}$} &\colhead{$\alpha_{2}$} & \colhead{L$_{b}$ [10$^{36}$ erg s$^{-1}$]} &\colhead{$C$} & \colhead{$C_{crit}$}}
\startdata
4-25~keV & 0.88 \left(1.22^{+0.22}_{-0.23}\right) & 2.28 \left(2.15^{+0.58}_{-0.43}\right) & 21.54 \left(28.20^{+11.87}_{-\phn8.73}\right) & 76.76 & 168.59 \\
12--25~keV & 0.39 \left(1.72^{+0.38}_{-0.48}\right) & 2.89 \left(2.49^{+0.53}_{-0.38}\right) & \phn6.80 \left(\phn7.09^{+\phn1.47}_{-\phn2.02}\right) & 38.91 & 125.51
\enddata
\tablenotetext{a}{Parameters are reported as the best fit value ($50\%\pm34\%$), for comparison with our disk XLF parameters in Table \ref{tab:XLFpars}}
\end{deluxetable*}

\bibliography{biblio}{}
\bibliographystyle{aasjournal}

\end{document}